\documentclass{aa}  
\usepackage{natbib}
\bibpunct{(}{)}{;}{a}{}{,} 

\usepackage{graphicx}

\usepackage{txfonts}
\usepackage{}

\newcommand{\Gaia}{Gaia\,}

\begin{document}

   \title{Accurate photometric calibration by fitting the system transmission}

   \author{S. Garrappa
          \inst{1}
          \and
          E. O. Ofek\inst{1}
          \and
          S.~Ben-Ami\inst{1}
          \and
          D.~Polishook\inst{1}
          \and
          A.~Gal-Yam\inst{1}
          \and
          Y.~Shvartzvald\inst{1}
          \and
          A.~Krassilchtchikov\inst{1}
          \and
          R. Konno\inst{1}
          \and
          E.~Segre\inst{1}
          \and
          Y. M.~Shani \inst{1}
          \and
          Y.~Sofer-Rimalt\inst{1}
          \and 
          M.~Engel\inst{1}
          \and
          A. Blumenzweig\inst{1}          
          }

   \institute{Department of Particle Physics and Astrophysics, Weizmann Institute of Science, 76100 Rehovot, Israel.\\
              \email{simone.garrappa@weizmann.ac.il}\\
              \email{eran.ofek@weizmann.ac.il}
             }

   \date{Received 13 December 2024 / Accepted 9 May 2025}

  \abstract 
   {Transforming the instrumental photometry of ground-based telescopes into a calibrated physical flux in a well-defined passband is a major challenge in astronomy. Along with the intrinsic instrumental difference between telescopes sharing the same filter, the effective transmission is continuously modified by the effects of the variable atmosphere of the Earth.}
   {We have developed a new approach to the absolute photometric calibration (i.e., tied to the CALSPEC standards) that simultaneously treats instrumental and atmospheric effects on an image-by-image basis by fitting the system transmission. This approach aims at breaking the 1\% absolute photometric accuracy which limits current calibration methods for ground-based observatories.}
   {We fit the transmission, as a function of wavelength, for each image. The fit is done by comparing the instrumental fluxes of stars in the image to the synthetic photometry of the stars given their spectrum and the transmission function which have free parameters. A key element that enables this approach is the set of about 220 million low-resolution spectra measured by \Gaia, which provides a large number of stellar calibrators in the image that are calibrated against the CALSPEC scale.}
   {We demonstrate the method using data from the Large Array Survey Telescope (LAST). We show that the residuals between observations and synthetic photometry of the \Gaia spectra in the fitted transmission
   have a standard deviation $<$1\% on an image-by-image basis, with no spatial and color dependencies. The median accuracy of the zero-point throughout the image is between 3-5 mmag, depending on the total image exposure. Furthermore we show that this method provides high stability over long temporal scales.}
   {}


   \maketitle
%
%-------------------------------------------------------------------

\section{Introduction}

A major challenge in astronomical observations is to accurately transform the instrumental photometry into a calibrated flux in a well-defined passband. Such calibration is typically required for either measuring the flux of the source (in physical units), or to make a comparison between the flux measured by different telescopes.
Ground-based surveys regularly provide an absolute calibration of their photometry to the 1-2\% level (e.g., \citealt{2008ApJ...674.1217P}, \citealt{Ofek+2012_photCalib}, \citealt{2012PASP..124..854O}, \citealt{2019PASP..131a8003M}, \citealt{2020ApJS..251....6M}).
However, breaking this $\sim$1$\%$ barrier is difficult.
Furthermore, the photometry based on standard methods is not provided in a well defined band.
One limiting factor is the intrinsic instrumental difference between astronomical facilities. Even when the same filter is used by different telescopes, they do not necessarily share the same optics, coatings, and detectors, which may modify slightly the effective transmission. Furthermore, when comparing the flux taken with the same filter using two telescopes, their effective transmission may be different due to e.g., local atmosphere. Furthermore, the variable atmosphere constantly changes the effective transmission, resulting in flux variations even if the source is constant. All these complications impose the historical limits on the possibility of achieving absolute calibrations with an accuracy of less than a few percent (see \citealt{2006ApJ...646.1436S} for a review).

The classical photometric calibration method was to observe a set of standard stars at different air mass, and to use this to measure the system photometric zero point as a function of air mass (e.g.,using Landolt stars; \citealt{1992AJ....104..340L}). The main limitation of this approach is the simplistic assumption that the zero point depends only on air mass (i.e., that atmospheric transmission is otherwise uniform). A better approach would be to have a set of standard stars with a density high enough such that some will be present in any random observation (e.g., \citealt{Skrutskie+2006_2MASS,Ofek2008_TychoMagCalib,Ofek+2012_PTF_photCat,2020ApJS..251....6M}).

A powerful approach for photometric calibration adopted by several sky surveys is the $\ddot{u}$bercalibration method (\citealt{2008ApJ...674.1217P}). In this approach the photometry is first calibrated internally by using stars that appear in two, somewhat overlapping, images defining a self-consistent system for instrumental magnitudes. After that, an external calibration is performed using a set of photometric standard stars that serve as calibrators with accurate measurements of their spectral energy distribution (SED). A recent application of this calibration approach is the one used for the \Gaia mission \citep{2016A&A...595A...1G}, for which the photometric G band has been defined \citep{2016A&A...595A...7C}. Indeed in the case of \Gaia, its photometry, as well as spectrophotometric measurments are consistently calibrated to the level of a few mmag \citep{2016A&A...595A...7C,2023A&A...674A...2D}. However, for ground-based observations, the temporal variations in the effective passbands are limiting the accuracy of this method. 

A conceptual solution to this problem is to somehow measure the atmospheric and system transmission for each image obtained by the telescope. Previous works have discussed the importance of having a precise description of the time-dependent optical transmission of the atmosphere by e.g., realtime measures on site (e.g., \citealt{2010ApJ...720..811B}, \citealt{2018AJ....155...41B}, also \citealt{2007PASP..119.1163S} for a comprehensive overview). \cite{2010ApJ...720..811B} shows how auxiliary spectroscopic measurements of a set of calibrator stars during the night allow to characterize the atmospheric absorption of different atmospheric constituents and their variation. However, auxiliary spectroscopic measurements on site require dedicated instruments to monitor the fields observed by the main survey (e.g., the Auxiliary calibration telescope of the Vera C. Rubin Observatory, \citealt{2009arXiv0912.0201L})

The method presented in this work consists of fitting the system transmission for each image by combining parametric models for the instrumental transmission and for the atmospheric transmission, fitted using a large number of stellar calibrators across the field of view. The method aims to obtain high accuracy and stability of the photometry over time, and to break the 1$\%$ barrier of the current state-of-the-art calibrations for ground-based facilities. Critical to this work is the existence of accurately calibrated low-resolution spectra of a large number of stars over the entire sky. Therefore, this work became feasible only after the appearance of the \Gaia catalog of $\sim$220 million low-resolution spectra \citep{2023A&A...674A...2D}, that provide a sufficient number of diverse calibrators in wide-field images. A similar approach was suggested by \citealt{2023A&A...674A...3M}, where the instrument model for the \Gaia BP and RP spectrophotometers was derived by fitting correction terms to the nominal, pre-launch passband of the instrument. In our approach, we also make use of synthetic photometry to match the predicted and observed counts in our detector for each source, following the technique outlined in \citealt{2012PASP..124..140B}. Following the description of the methodology, we explore the effects of atmospheric transmission variations on the photometry.
We demonstrate the suggested method using data obtained by the Large Array Survey Telescope (LAST; \citealt{2023PASP..135f5001O}, \citealt{2023PASP..135h5002B}) and show that it can provide stable and consistent absolute photometric calibration to a level of 3-5 mmag. Throughout this work, the term absolute calibration refers to the process of determining a solution for the system’s effective transmission. Since the Gaia spectra used here are calibrated against the CALSPEC scale \citep{2021MNRAS.501.2848A,2023A&A...674A...3M}, the resulting calibration is inherently tied to the CALSPEC standards \citep{2014PASP..126..711B}.

In Sec. \ref{sec:problem} we present our proposed solution to this problem. In Sec. \ref{sec:transmission_basis} we dissect the transmission model into all of its components. In Sec \ref{sec:method} we describe the approach to the model optimization based on synthetic photometry, and in Sec. \ref{sec:sensitivity_atm} we evaluate the sensitivity of our model to variations of the atmospheric parameters. In Sec. \ref{sec:real_data_LAST} we apply the method on real data from LAST, while in Sec. \ref{sec:results} we discuss the results. In Sec. \ref{sec:software} we describe the code that we used to implement the method and in Sec. \ref{sec:summary} we summarize the work and discuss future efforts. In Appendix \ref{app:header_keys} we suggest a standard for FITS headers to store the photometric transmission parameters.

\section{Description of the methodology}
\label{sec:problem}
The problem of deriving the throughput as a function of the wavelength $\lambda$, given a set of calibrators with a known spectrum, can be generally formulated as follows:
\begin{equation}
    F_{inst,j} = \sum_{i}^{N} \alpha_{i}T_{i}(\lambda)S_{j}(\lambda)\,.
    \label{eq:F_T}
\end{equation}
Here F$_{inst,j}$ is the observed instrumental flux of the $j$-$th$ calibrator, $S_{j}$($\lambda$) is its known spectral flux at the wavelength $\lambda$, $T_{i}(\lambda)$ is a set of $N$ positive-definite function-basis that span the spectral space.
Finally, $\alpha_{i}$ are the free parameters we would like to find, where the system transmission is given by
\begin{equation}
    T(\lambda) = \sum_{i}^{N}{\alpha_{i}T_{i}(\lambda)}.
    \label{eq:T}
\end{equation}
The flux and spectra in this problem can be generally written in physical (calibrated) units. In this work, we convert them to instrumental photon counts only when required (e.g., Sec. \ref{sec:method}). A necessary condition for solving Equation~\ref{eq:F_T} is that the spectra of the calibrators $S_{j}(\lambda)$ will be diverse enough such that finding $\alpha_{i}$ will be possible.

A straightforward approach to solve for $\alpha_{i}$ is to assume the $T_{i}(\lambda)$ are top-hat functions, and in this case, the problem can be solved using linear least square or positive linear least square methods.
To validate the feasibility of this approach we have performed the following simulations. We selected 1000 random \Gaia spectra, generated a transmission curve, and calculated the synthetic flux of the \Gaia spectra using this transmission. Next, we added noise to the synthetic fluxes (to mimic $F_{inst, j}$) and attempted to solve Equation~\ref{eq:F_T}.
Assuming $T_{i}$ is a set of top-hat functions we were able to recover $\alpha_{i}$ and the input $T(\lambda)$ only when the photometric relative errors were unrealistically small ($\lesssim10^{-3}$).
This indicates that the problem requires some regularization.
A reasonable regularization is to assume that the transmission function we are trying to recover is smooth, or composed of physical components.
For example, when we redid the simulations using $T_{i}$ which are polynomials or other functions (see below) that problem became solvable even in the presence of realistic relative noise (e.g., $\sim$0.1).

There are many options for choosing a base of $T_{i}$, which are likely to work.
Here, we decided to use a set of functions which are physically motivated, but to add freedom via low-order perturbing functions. In addition, atmospheric absorption can be attributed to several known mechanisms with some free parameters. In the next section we describe the basis $T_{i}$ used in this work.
The proposed basis contains a large number of free parameters.
However, by investigating the effect of each parameter we select a subset of parameters that have the largest effect on the results and describe the transmission to acceptable accuracy.

\section{The transmission model basis}\label{sec:transmission_basis}
The model for the overall transmission includes a series of different components such as the instrumental transmittance of the telescope optical elements (T$_{tel}$), the quantum efficiency (QE) of the camera (T$_{det}$), the atmospheric transmission (T$_{atm}$), and a component that introduces additional perturbations in the model (T$_{pert}$). The model can be described as follows:

\begin{equation}\label{eq:transmission_model_ota_atm}
    T(\lambda) = T_{tel}(\lambda)T_{det}(\lambda)T_{atm}(\lambda)T_{pert}(\lambda).
\end{equation}
Each of these components, is composed of several independent sub-components that we present in details below. Some of these components may be known and other may have some free parameters.

\subsection{Telescope and detector}\label{sec:OTA_transmission}
The first two terms of Equation \ref{eq:transmission_model_ota_atm}, T$_{tel}(\lambda)$ and T$_{det}(\lambda)$, describe the known transmission of the different instrumental components of our system. The first one describes the components of the telescope itself, and can be in turn described as the factorization of the transmissions of the different components (T$_{C_{i}}$) within the optical path in the telescope:

\begin{equation}\label{eq:T_tel}
    T_{tel}(\lambda) = \prod_{i}^{N}{T_{C_{i}}} .
\end{equation}
As for T$_{det}(\lambda)$ is usually the product of the quantum efficiency (T$_{QE}(\lambda)$) of the camera installed on the telescope, as typically obtained by laboratory measurements and the transmission of the filter T$_{F}(\lambda)$ used for observations

\begin{equation}\label{eq:T_det}
    T_{det}(\lambda) = T_{QE}(\lambda)T_{F}(\lambda) .
\end{equation}

\subsection{Perturbations}\label{sec:perturbations}

In order to introduce some freedom in the shape of the instrumental transmissions measured in the laboratory, one can consider adding perturbations to the model with a number of free parameters. This is very important since even small changes in the transmission can lead to measurable differences in the flux estimation of sources (see Appendix \ref{app:appendix_a}). One possible way to add such perturbations, is to use a base of low-order (N) polynomial functions described by the general form

\begin{equation}\label{eq:pert}
    T_{pert}(\lambda) =  \sum_{n=0}^{N} c_{n}P_{n}(\lambda), 
\end{equation}
where $c_{n}$ are the fitted coefficients and $P_{n}$ is the n-th order polynomial. A similar perturbation approach is applied in \citealt{2023A&A...674A...3M}, where a basis of Legendre polynomials is used and the perturbation is the exponential function of the linear combination of the basis, to guarantee the non-negativity of the perturbations.

Another approach, which we use in this work, is to replace the measurement of T$_{QE}(\lambda)$ of the camera with an analytic function with a similar shape, but with a number of free parameters that allow adjusting its shape. For example, in this work we replace T$_{det}$ with a skewed Gaussian function multiplied by a polynomial basis

\begin{equation}\label{eq:general_Tdet_model}
    T_{det} = f(\lambda;A,\mu,\sigma,\gamma) exp\left(  \sum_{n=0}^{N} c_{n}P_{n}(\lambda) \right),
\end{equation}
where the polynomial basis is treated as argument of the exponential function in order to have non-negative values for the transmission model. The skewed Gaussian function $f(\lambda;A,\mu,\sigma,\gamma)$ is expressed as

\begin{equation}\label{eq:skewed_gaussian}
f(\lambda;A,\mu,\sigma,\gamma) =
    \frac{A}{\sigma\sqrt{2\pi}}e^{[-(\lambda-\mu)^{2}/2\sigma^{2}]}\biggl\{1 + erf\Bigl[ \frac{\gamma(\lambda-\mu)}{\sigma\sqrt{2}} \Bigr] \biggr\},
\end{equation}

where A is the amplitude, $\mu$ is the center, $\sigma$ is the equivalent width and $\gamma$ is the skeweness. Such modeling allows to introduce modifications in the shape of the total effective transmission introduced by the optics and other external factors. This method is useful, for example, when the actual QE curve of the detector is similar to a skewed Gaussian function. In Appendix \ref{app:appendix_b} we explore the effects of this choice on the results.\\

\subsection{Atmospheric transmission}
\label{sec:atmospheric_transmission}
The contribution of the atmosphere to the overall transmission is, in most cases, the major factor that limits the accuracy of flux measurements to the 1--2\% level. The atmospheric effects are of two different kinds: scattering and absorption. The former includes Rayleigh scattering of light by molecules of gases in the atmosphere, and Mie scattering by aerosol particles. All remaining processes are due to absorption of light by atmospheric constituents (gases of different species). In this work, we use analytical and empirical parameterizations for the atmospheric components and the relative air mass of the constituents from the SMARTS v2.9.8 model (see \citealt{2019SoEn..187..233G} and references therein).

\begin{figure*}[ht!]
\centering
\includegraphics[width=17cm]{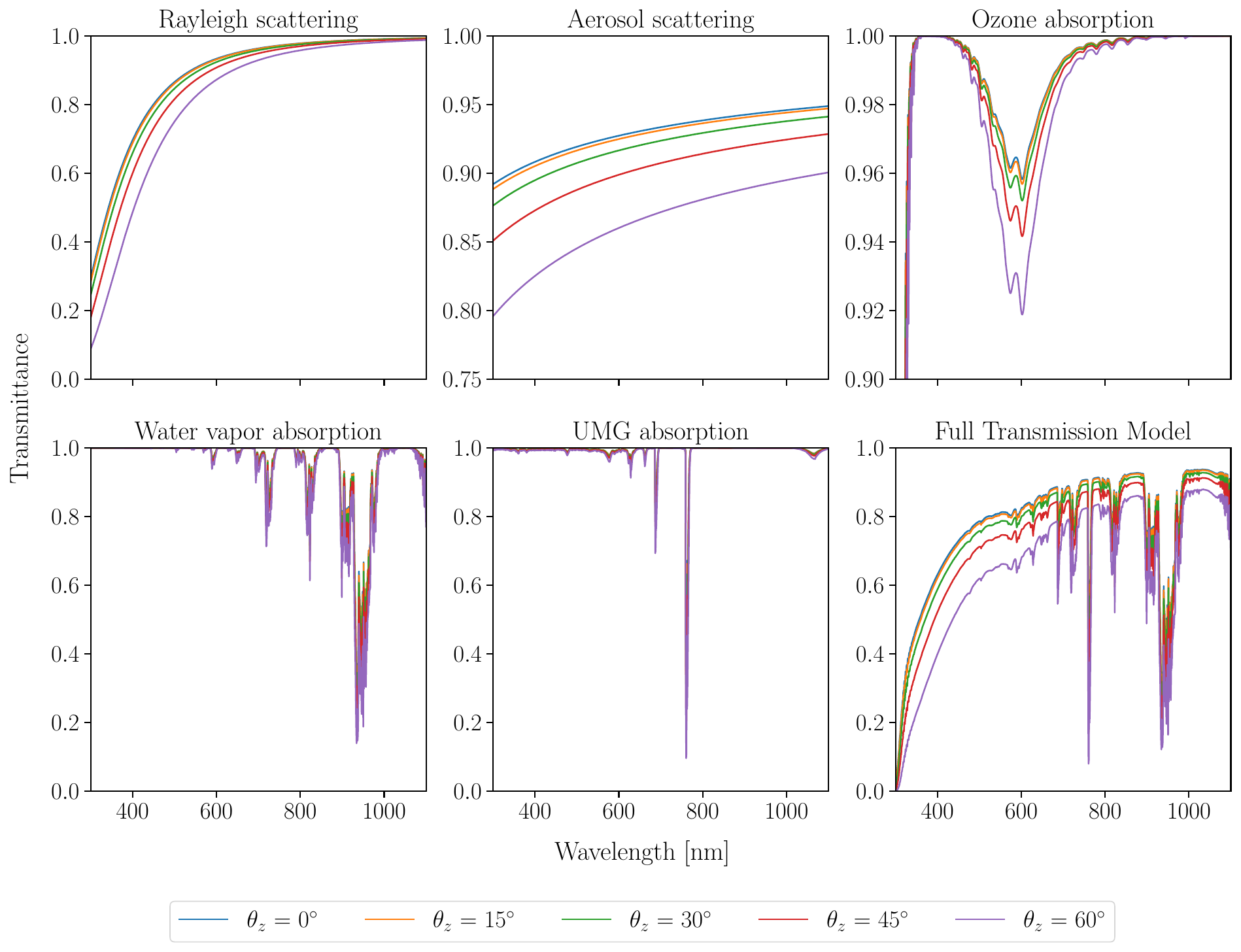}
\caption{Transmittance of each atmospheric component at different zenith angles. The panels in the top row show from left to right the transmittance for Rayleigh scattering, aerosol scattering and ozone absorption. In the bottom panels, left to right panels are the transmittance of the absorption by water vapor, absorption by the uniform mixture of gases, and the product of all atmospheric components.
\label{fig:ATM_Models}}
\end{figure*}

The total transmission due to light traveling throughout the atmosphere can be computed as the product of the transmissions from each of the N (=5) processes considered:

\begin{equation}
    T_{atm}(\lambda) = \prod^{i = N}_{i = 1} T_{atm,i}(\lambda).
\end{equation}
Each T$_{atm,i}$ is usually described by the analytical law

\begin{equation}
    T_{atm,i}(\lambda) = e^{-m_{i}\tau_{i}(\lambda;\hat{\mathbf{\gamma}})},
\end{equation}
where $m_{i}$ is the constituent's  air mass over the column at a fixed zenith angle $\theta_{z}$ given by

\begin{equation}\label{eq:eff_airmass}
    m_{X} = \frac{1}{\cos(\theta_{z}) + \eta_{0}*\theta_{z}^{\eta_{1}}(\eta_{2} - \theta_{z})^{\eta_{3}}}.
\end{equation}
Here $X$ defines the component, and $\eta_{1}$, $\eta_{2}$ and $\eta_{3}$ are parameterizations in \citealt{2019SoEn..187..233G} to estimate the effective airmass of each atmospheric component. We list in Appendix \ref{app:airmass_pars} the parameters from SMART v2.9.8. The term $\tau_{i}(\lambda;\hat{\mathbf{\gamma}})$ is the constituent optical depth at a specific wavelength $\lambda$ per unit air mass, and $\hat{\mathbf{\gamma}}$ is a set of additional parameters. In our transmission model, we consider a total of five atmospheric components: Rayleigh scattering, aerosol scattering, ozone absorption, water vapour absorption, and absorption from a uniform mixture of gases. The respective contribution of these components is discussed below and it is shown in Figure \ref{fig:ATM_Models}.

Rayleigh scattering: describes the scattering of light by gas molecules. The optical depth of this process can be described by a fourth degree polynomial of the wavelength $\lambda$ and is parameterized by average empirical results in the literature (SMART v2.9.8 uses results from \citealt{Bodhaine1999}, \citealt{Bucholtz1995} and \citealt{Tomasi2005})

\begin{equation}
    \tau_{R,\lambda} = \frac{p}{p_{0}}(117.3405\lambda^{4} - 1.5107\lambda^{2} +
    0.017535 - 0.00087743\lambda^{-2})^{-1},
\end{equation}
where $p_{0} = 1013.25\,$mbar is the standard pressure, $p$ the ground-level pressure (in mbar), and $\lambda$ is the wavelength measured in $\mu$m. The effective airmass of the component determining the Rayleigh scattering is calculated from Equation \ref{eq:eff_airmass} using the values of the $\eta$ parameters in Table \ref{tab:airmass_parameters}. The only free parameter for this component is the surface pressure $p$. The top-left panel of Figure \ref{fig:ATM_Models} shows the transmission profile from Rayleigh scattering at different air mass, using $p = 965\,$ mbar\footnote{This is the average atmospheric pressure at the LAST site.}.

Aerosol scattering: describes the interaction with aerosol mixtures in the atmosphere. The analytical formulation for the optical depth of this component (also known as $\mathrm{\AA}$ngstrom's law) is

\begin{equation}
    \tau_{a,\lambda} = \tau_{a500}(2\lambda)^{-\alpha},
\end{equation}
where $\lambda$ is measured in $\mu$m, $\tau_{a500}$ is the aerosol optical depth (AOD) at 500\,nm and $\alpha$ is the $\mathrm{\AA}$ngstrom exponent that depends on the average grain size of the aerosol mixture. Therefore, this component has two free parameters. In the typical conditions of rural aerosol model $\tau_{a500}$ $\approx$ 0.084, while in desert-dust aerosol conditions it gets up to $\tau_{a500}$ $\approx$ 0.54. In desert-like conditions, $\alpha$ is constant through the whole wavelength range of interest in our work, and has typical value $\approx$0.6 in moderate presence of dust \citep{2019SoEn..187..233G}. The upper-central panel of Figure \ref{fig:ATM_Models} shows the transmission profile from aerosol scattering at different air mass, using $\tau_{a500}$ = 0.084 and $\alpha$ = 0.6\,. The effective airmass of the component determining the aerosol scattering is calculated from Equation \ref{eq:eff_airmass} using the values of the $\eta$ parameters in Table \ref{tab:airmass_parameters}. 

Ozone absorption: is the absorption due to the total O$_{3}$ column in the atmosphere. Absorption from O$_{3}$ suppresses severely the transmission in the near-UV band below about 360 nm, and its optical depth is described by the Beer–Lambert–Bouguer law 

\begin{equation}
    \tau_{o,\lambda} = u_{o}A_{o,\lambda},
\end{equation}
where u$_{0}$ is the length of the total O$_{3}$ column expressed in Dobson Units (1\,DU = 10 $\mu$m-atm) and A$_{o\lambda}$ is the known spectral absorption of O$_{3}$. Tabulated values for A$_{o\lambda}$ are taken from SMARTS v2.9.8 \citep{2019SoEn..187..233G}. The parameter u$_{0}$ is the only parameter that can be assumed free in this component. The upper-right panel of Figure \ref{fig:ATM_Models} shows the transmission profile from O$_{3}$ absorption at different air mass, assuming u$_{0} = 300$\,DU. Other than the severe absorption below $\sim$360\,nm mentioned above, the transmission profile shows another absorption feature in the Chappuis band (500\,nm $<$ $\lambda$ $<$ 700\,nm) that accounts for up to $\sim$8\% attenuation at 600\,nm for moderate air mass. The effective airmass of the O$_{3}$ column is calculated from Equation \ref{eq:eff_airmass} using the values of the $\eta$ parameters in Table \ref{tab:airmass_parameters}.

Water vapor absorption: is the absorption due to the water vapor in the atmosphere. Unlike the previous components, its transmission has a different functional dependence with the air mass, and it can be described as \citep{2001SoEn...71..325G}:

\begin{equation}
    T_{v,\lambda} = e^{-[(m_{w}p_{w})^{1.05}f^{n(\lambda)}_{v}B_{v}(p)A_{v,\lambda}]^{c(\lambda)}}.
\end{equation}
Here m$_{w}$ is the air mass for the water vapor (calculated with Equation \ref{eq:eff_airmass}, using the values of $\eta$ for H$_{2}$O in Table \ref{tab:airmass_parameters}), and p$_{w}$ the total amount of precipitable water (in cm). The scaling factor $f^{n(\lambda)}$ compensates pressure inhomogeneities along the water vapor column following the Curtis-Gordon approximation \citep{1989ApOpt..28.3792P} and $B_{v}(p)$ is a correction term dependent on the atmospheric pressure, that accounts for the different absorption intensity with respect to the distance from the band center. Lastly, $c$ and $n$ are wavelength-dependent exponents, which are tabulated in \citealt{2019SoEn..187..233G}. Apart from using the parameterizations given in \cite{2001SoEn...71..325G} and the effective water vapor air mass of the observation, the only free parameters in this component are the amount of precipitable water p$_{w}$ and the atmospheric pressure $p$. The value of p$_{w}$ depends strongly on the site location and conditions, with typical values between 1--2 cm. 

Uniform gas mixture absorption: is the absorption due to a uniform mixture of gases following the Standard US atmosphere model \citep{1992P&SS...40..553N}, to include remaining contributions not considered for the other components. Each gaseous component contributes to the total transmittance with the functional form given by \cite{2001SoEn...71..325G}:

\begin{equation}
    T_{g,\lambda} = e^{-(m_{g}u_{g}A_{g,\lambda})^{a}}.
\end{equation}
Here m$_{g}$ is the gas mixture air mass, calculated with Equation \ref{eq:eff_airmass} and the values of the parameters $\eta$ from Table \ref{tab:airmass_parameters} of all the gaseous components from O$_{2}$ to NH$_{3}$ weighted by their respective abundance in the Standard US atmosphere model. The term u$_{g}$ is the altitude-dependent gaseous pathlength and A$_{g,\lambda}$ is the spectral absorption coefficent. The tabuled values of u$_{g}$ and A$_{g,\lambda}$, as well as the values of the exponent $a$ for different wavelength ranges, are the same as implemented in the SMARTS v2.9.8 model (\citealt{2019SoEn..187..233G}). The free parameters of this component are the surface pressure and temperature.

The transmittance of each atmospheric component included in our model is shown in the subpanels of Figure \ref{fig:ATM_Models}. Each transmittance is presented at zenith angles ($\theta_{z}$) of 0$^{\circ}$, 15$^{\circ}$, 30$^{\circ}$, 45$^{\circ}$ and 60$^{\circ}$. The bottom-right panel of Figure \ref{fig:ATM_Models} shows the resulting transmittance after multiplying all the single components.

\section{Fitting procedure}\label{sec:method}
We outline here the method for the optimization of the overall transmission model described in the previous sections. This is obtained by comparing the synthetic photometry with the modeled transmission of sources with calibrated spectra with the instrumental photometry (photon counts) measured from the images. An overview of synthetic photometry can be found in \citealt{2012PASP..124..140B}, while here we describe the elements we adopt in our procedure.

Fluxes measured in instrumental photometry are usually reported in total number of photons within a fixed photometric aperture or point-spread function (PSF) fitting. We define here as N$_{e}$ the rate of measured electrons (in $e^{-}/s$) obtained by dividing the integral photon flux by the image exposure time. For a source with spectral flux F($\lambda$) and a system with transmission T($\lambda$), the electron rate is given by

\begin{equation}\label{eq:countrate}
    N_{e,syn} = \frac{A_{g}}{hc}\int_{\lambda}F(\lambda)\lambda T(\lambda)d\lambda,
\end{equation}
where A$_{g}$ is the geometrical collecting area of the telescope, $h$ is the Planck constant and $c$ the speed of light. In the integral term of Equation \ref{eq:countrate}, the only term containing free parameters is T($\lambda$), defined in the most general form in Equation \ref{eq:transmission_model_ota_atm}. The optimization to find the set of best-fit parameters ($\hat{\mathbf{\varrho}}$) for the transmission T($\lambda;\hat{\mathbf{\varrho}}$) consists then in minimizing the residuals between the measured instrumental electron rate and the predicted electron rate from synthetic photometry:

\begin{equation}\label{eq:residuals_wo_fieldcorr}
    R = N_{e,obs} - N_{e,syn}. 
\end{equation}

So far we assumed that the transmission is constant across the field of view. This is generally not accurate, as differences can be caused by factors like non-uniform illumination or response of the sensor, or residuals from dark and flat field corrections (e.g., \citealt{Padmanabhan+2008_SDSS_ImprovedPhotometricCalibration, Ofek+2012_photCalib}). Another kind of non-uniformity across the field of view can be introduced by the presence of cloud structures. This is known as gray extinction (as it is wavelength-independent) and can introduce significant spatial structures in the zero-point distribution across the field of view (e.g., \citealt{2014AJ....147...19B}).  We therefore add a polynomial term to Equation \ref{eq:residuals_wo_fieldcorr} to correct for these non-uniformities

\begin{equation}\label{eq:polynomial_field_corr}
\begin{split}
    P(x,y) = \sum_{i=1}^{n_{x}}\sum_{j=1}^{n_{y}}k_{i,j}P_{i}(x)P_{j}(y),
\end{split}
\end{equation}
where P$_{i}(x)$ and P$_{j}(y)$ are a basis of low-order ($n$) polynomials, $x$ and $y$ the pixel coordinates of the sensor, and $k_{i,j}$ the fitted coefficients. It is convenient to minimize these residuals in logarithmic space, therefore the field corrections P(x,y) can be expressed in magnitudes and the formulation of the residuals to be minimized becomes 

\begin{equation}\label{eq:residuals_w_fieldcorr}
\overline{R} = 2.5log_{10}\left(\frac{A_{g}C}{hcN_{e,obs}}\int_{\lambda}F(\lambda)\lambda T(\lambda;\hat{\varrho})d\lambda\right) + P(x,y;\hat{k}),
\end{equation}
where $\hat{\mathbf{\varrho}}$ is the set of free parameters in the transmission model, $\hat{k}$ is a set of free parameters in the spatial polynomial corrections (Equation \ref{eq:polynomial_field_corr}) and $C$ is a normalization factor that corrects for the effective collecting area of the system. $\overline{R}$ is now in magnitude space.\\

\section{Sensitivity to atmospheric parameters}\label{sec:sensitivity_atm}

\begin{figure*}[ht!]
\centering

\includegraphics[width=17cm]{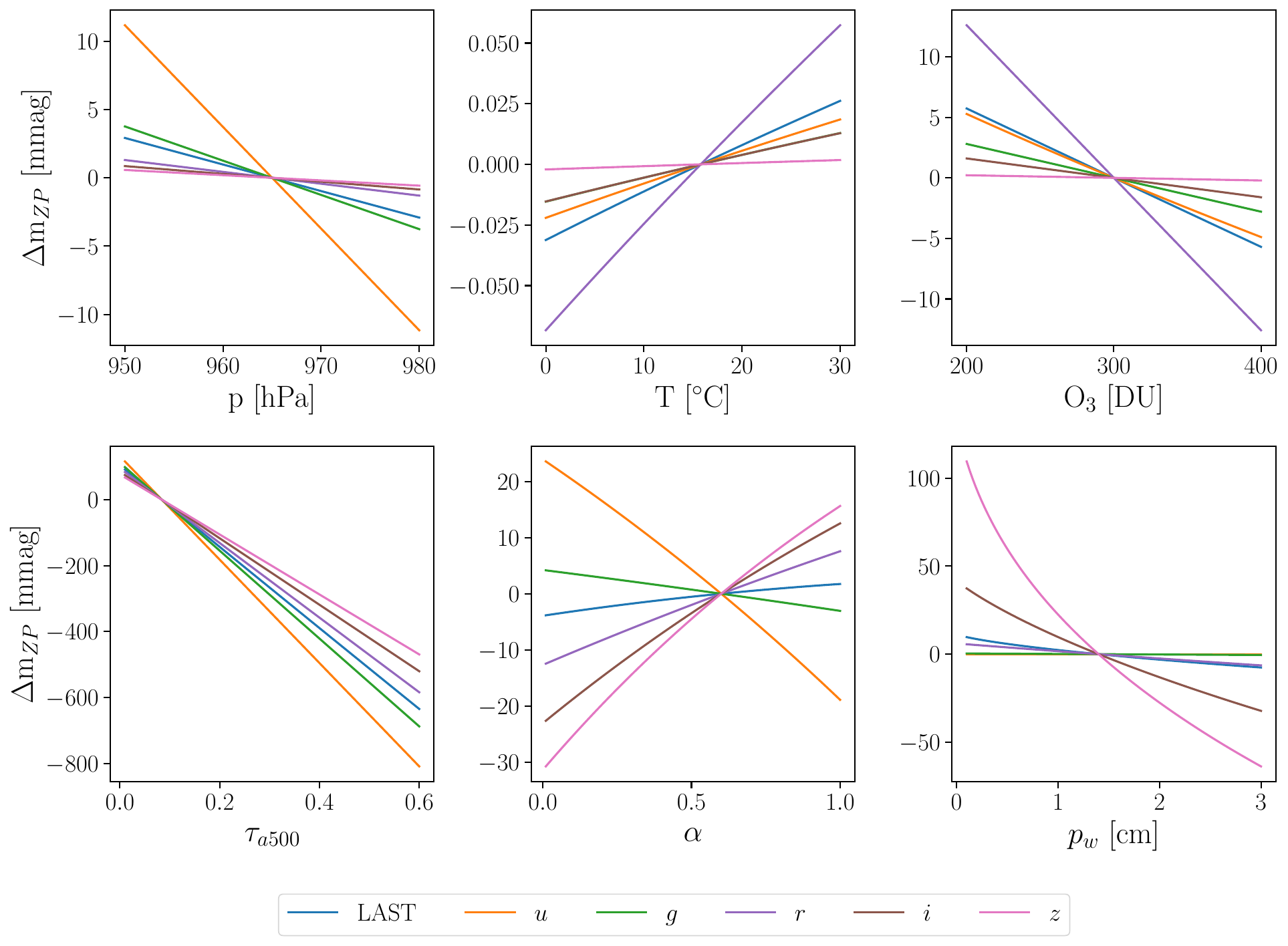}

\caption{Change in zero point magnitude in selected bands as a function of each atmospheric parameter. The data points are obtained by freeing the atmospheric parameter of interest in the transmission model and calculating the variation of the zero point as a function of the parameter. The reference zero point for each curve is calculated for the atmospheric values in Table \ref{tab:model_parameters} and T = 15.8 $^{\circ}$C.
\label{fig:deltamag_atm}}
\end{figure*}

The atmospheric component of our transmission model has a total of six free parameters ($p, \tau_{a500}, \alpha, u_{0}, p_{w}$ and $T$) that were described in Sect. \ref{sec:atmospheric_transmission}. Since some of these parameters are degenerate, and most of them are not expected to vary significantly outside of a physically-motivated range, it is useful to investigate the sensitivity of the photometry to the variation of the free parameters. Hence, we assume a mock broadband transmission model and we calculate the variation of a putative zero-point of the system (hence the integrated transmission) for variations of these parameters within typical ranges. In the panels of Figure \ref{fig:deltamag_atm} we show the variation (in magnitudes) of the zero point for selected bands by varying each time a single atmospheric parameter while keeping the rest of the parameters at their nominal values. We make this exercise on the LAST passband (introduced in Sec. \ref{sec:real_data_LAST}) and on the u,g,r,i,z SDSS filters \citep{2010AJ....139.1628D}. For most of the parameters, there is a linear dependency to the induced variation in the zero point. Two exceptions are the parameters $\alpha$ and $p_{w}$, which introduce non-linear variations on large ranges. 

For the LAST band, the parameter $\alpha$ from the aerosol component introduces variations $<$1$\%$ even at the edges of the range considered, and $p_{w}$ is supposed to vary between 1--2 cm on an average day, which will still introduce variations $<$1$\%$ in the most extreme cases. Moreover, the behavior of both parameters can be considered fairly linear across small intervals within the fiducial range. The plots show that differences between the real and assumed/measured value for the other atmospheric parameters will introduce an error on the zero-point of $<$\,1$\%$ in the LAST band, with the exception of $\tau_{a500}$ that can lead to errors of the order of $\sim$10$\%$. From this exercise, we can identify two parameters with contributions to the photometric error in the LAST band that could exceed the 5 mmag level which are the aerosol $\tau_{a500}$ and the amount of precipitable water $p_{w}$. Therefore, in principle one can consider keeping these parameters free in the model when applied to LAST data. 

It is important to remark that the previous considerations are for the specific case of the LAST broad passband. The plots in Figure \ref{fig:deltamag_atm} show how the atmospheric components can affect very differently each single narrower bands (see, e.g., Figure \ref{fig:ATM_Models}. The contribution of each parameter to the photometric error depends on the specific band considered, e.g., the $u$ band is more sensitive to Rayleigh scattering and the aerosol $\alpha$ parameter, while the precipitable amount of water $p_{w}$ dominates the contribution to the $z$ band. Therefore we can conclude that, in our approach, the choice of the subset of free atmospheric parameters will depend on the specific band to be calibrated.  

\begin{figure}[ht!]
\begin{center}

\includegraphics[width=\linewidth]{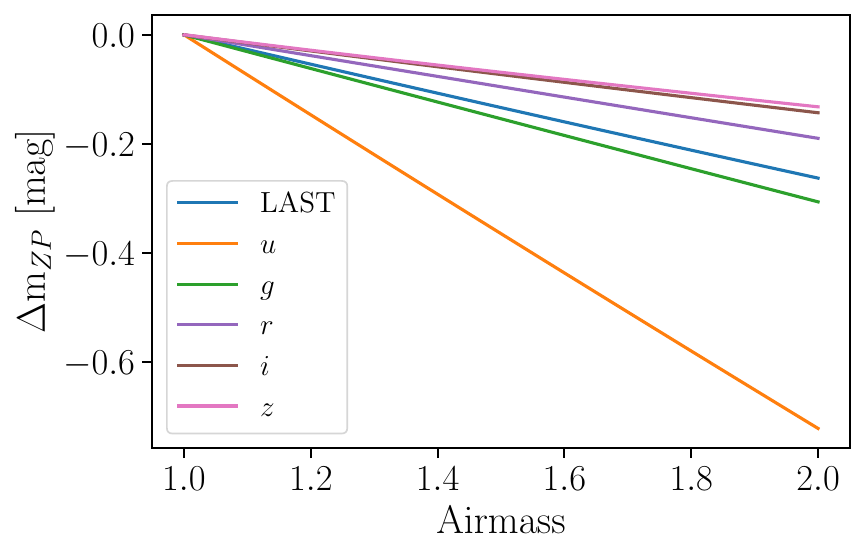}\\

\end{center}

\caption{Change in zero point magnitude in selected bands as a function of the air mass. The data points are obtained by freeing the air mass while keeping each transmission fixed. 
\label{fig:deltamag_airmass}}
\end{figure}

We repeat the same exercise for variations in air mass, in order to quantify the effect of uncertainties in the estimation of the airmass from the geometrical and empirical approximations used in the model. We select the same passbands used in the previous section and we calculate the variation of the zero point as a function of the air mass (Figure \ref{fig:deltamag_airmass}). The zero point has a linear dependency on the air mass, and for the case of the broad LAST band it varies by $\sim$0.3\,mag per unit air mass. Therefore, even assuming an uncertainty on the total effective air mass of the order of 1$\%$, the error in the photometric zero point is < 1$\%$ in the case of LAST. Among the selected bands, the SDSS $u$ filter shows the strongest dependency to air mass variations, with a slope of $\sim$0.7\,mag per unit air mass.

In addition to the uncertainties on the air mass, there is also the difference in the effective air mass in the different parts of the image (compared to the air mass calculated at the image center). This effect increases with the size of the image, and it is more prominent for observations made at large zenith angles. We find that the polynomial corrections in Equation \ref{eq:polynomial_field_corr} are effective in correcting also this non-uniformity. Alternatively, one can estimate the position dependent air mass for each star.

\section{Tests on real data from LAST}\label{sec:real_data_LAST}
Here we apply the method to data obtained with the Large Array Survey Telescope (LAST). We first give a brief description of the LAST system and the photometry pipeline. We then present the transmission model for the LAST system and the results of the method on a variety of images.  
\subsection{The LAST data and instrumental photometry}
LAST is an optical telescope system located at the Weizmann Observatory in the Israeli Negev desert. It is an array of 28 cm f/2.2 Rowe-Ackermann-Schmidt telescopes designed to study the variable and transient sky (\citealt{2020PASP..132l5004O},\citealt{2023PASP..135f5001O}, \citealt{2023PASP..135h5002B}). The full LAST array will consist of 72 telescopes (32 already operative), with each telescope providing a field of view of 3.3 deg $\times$ 2.2 deg with a plate scale of 1.25 arcsec/pixel on camera. The limiting magnitude (AB) of the system is 19.6 (21.0) in 20 s (20x20 s) exposures.

The application of the absolute photometric calibration method described in this section is based on the products of the LAST data pipeline described in \citealt{2023PASP..135l4502O}. The pipeline processes and calibrates 24 sub-divisions of each single image, and applies two different methods to extract the photometry for each source. The first method is aperture photometry with 3 different apertures (2, 4 and 6 pixel radius) and it is calculated summing all the pixels within each radius, after FFT-shifting the sources to the pixel origin. The second method is PSF-fit photometry, where the PSF is estimated for each sub-image and fit to all sources. As expected, the performance of the aperture photometry is better for bright sources (G $<$ 16 mag) and PSF photometry has better performance at the faint end. 
Since our procedure uses bright calibrators in the image, our default is to calibrate the aperture photometry (with 6 pixel radius). 

\subsection{The instrumental transmission of LAST}\label{sec:instr_transmission_LAST}
After introducing a general formulation for the instrumental transmission model in Sect. \ref{sec:OTA_transmission}, we now define each term specifically for the application to LAST data. In the case of LAST, the optical elements in each telescope are the primary mirror, the corrector plate and the field flattener lenses. Therefore, we rewrite Equation \ref{eq:T_tel} as

\begin{equation}
    T_{tel} = T_{XLT}T_{lenses},
\end{equation}
where $T_{XLT}$ is the mirror transmission with the Celestron Starbright XLT coating and $T_{lenses}$ the transmission of the corrector plate and field flattener. We consider the reflectivity of the primary mirror (T$_{XLT}$) and the transmission of the corrector plate (T$_{corr}$) from the  StarBright XLT coating technical datasheet\footnote{\url{https://www.celestron.com/pages/starbright-xlt-optical-coatings}}. The available measurements cover only a wavelength range that is smaller than the one where the LAST camera is sensitive. Therefore, these transmission curves are modelled with second-order polynomial functions and extrapolated to the desired range. Although measurements for the field flattener lens system are not available, we assume that its transmission has a similar wave dependence to that of the corrector.\\
The other instrumental term for the transmission is T$_{det}$ from Equation \ref{eq:T_det}. In the case of LAST, no filter is used and therefore
\begin{equation}
    T_{det} = T_{QE},
\end{equation}
where T$_{QE}$ is the quantum efficiency measured for the QHY600-PH CMOS camera mounted on each LAST telescope \citep{2023PASP..135f5001O}. Since the QE is dependent on the incidence angle, we expect the nominal QE to vary across the field of view.
In order to allow some optimization of the effective camera QE, we fit the laboratory measurements from \cite{2023PASP..135f5001O} with the skewed Gaussian function multiplied by a perturbation term (Equation \ref{eq:general_Tdet_model} and \ref{eq:skewed_gaussian}), with an 8th order Legendre polynomial basis. This procedure is done once and the polynomials parameters listed in Table \ref{tab:model_QE_parameters} are held fix during the calibration. We also fix the parameters $A$, $\gamma$ and $\sigma$, but we leave $\mu$ as a free parameter. In appendix \ref{app:appendix_c} we show the effects of this method across the field of view when this freedom is not included in the model, and color terms are observed in the final model residuals after the calibration.

\begin{figure}[ht!]
\includegraphics[width=\linewidth]{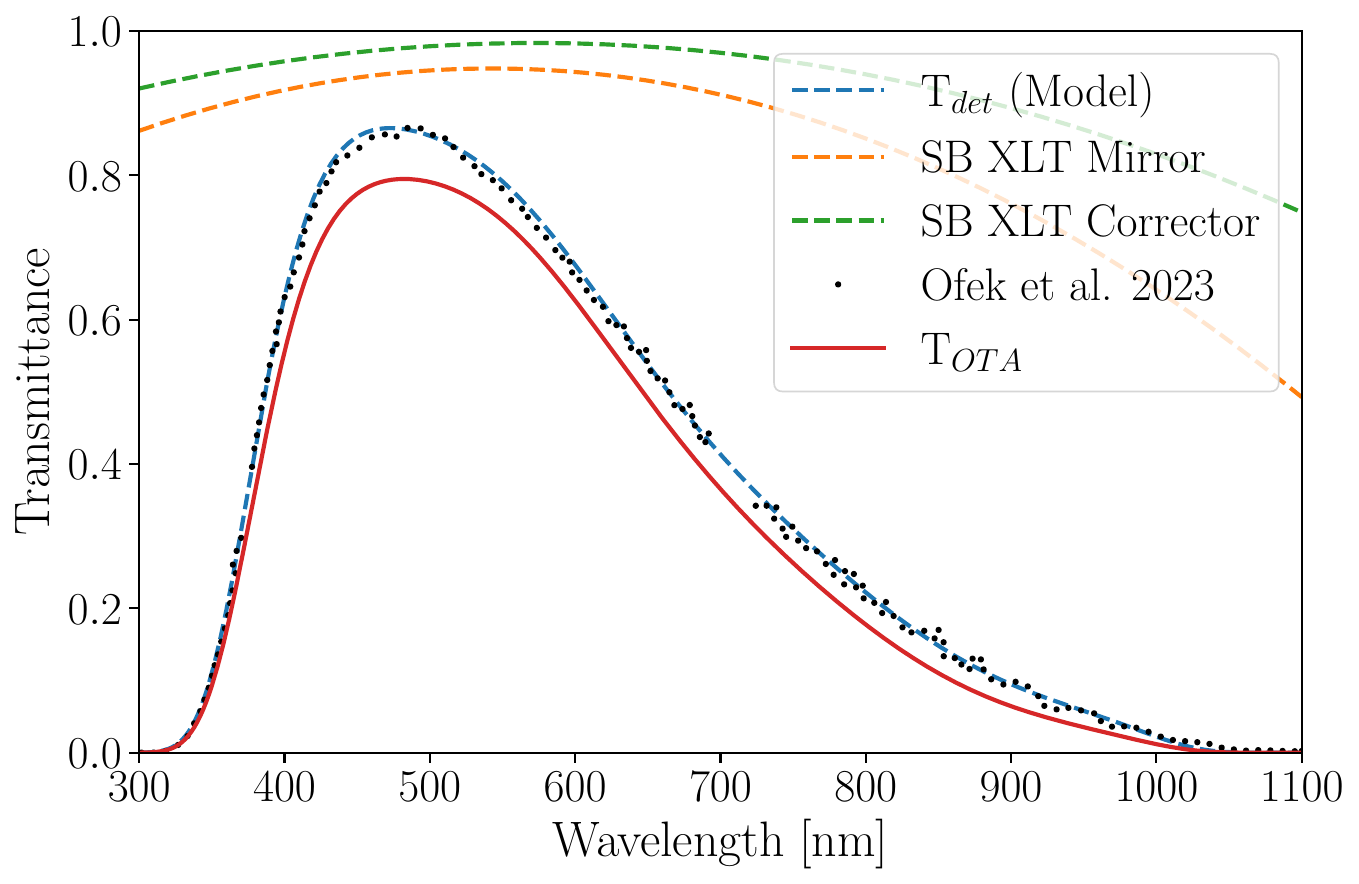}
\caption{Optical tube assembly instrumental transmission components. The black dots are the laboratory measurements from \citealt{2023PASP..135f5001O} of the QHY600PH QE. The dashed lines show the models of the single instrumental components, while the solid red line shows the total transmission given by the product of the single components.
\label{fig:ota_transmission}}
\end{figure}

To summarize, Figure \ref{fig:ota_transmission} shows each instrumental component of the transmission and the resulting shape $T_{OTA}$ (solid red line) given by 

\begin{equation}
    T_{OTA} = T_{tel}T_{det}.
\end{equation}

\begin{table}[h!]
\centering
\caption{Parameters for the T$_{OTA}$ of LAST.}
\begin{tabular}{lll}
\hline
        \textbf{Parameter Name} &                  \textbf{Component} &                \textbf{Value}  \\
\hline
A &             Skewed Gaussian &           328.19  \\
$\mu$ &             Skewed Gaussian &           570.97  \\
$\sigma$ &             Skewed Gaussian &           139.77  \\
$\gamma$ &             Skewed Gaussian &           -0.15  \\
$l_{0}$ &             Perturbation &           -0.30  \\
$l_{1}$ &             Perturbation &           0.34  \\
$l_{2}$ &             Perturbation &           -1.89  \\
$l_{3}$ &             Perturbation &           -0.82  \\
$l_{4}$ &             Perturbation &           -3.73  \\
$l_{5}$ &             Perturbation &           -0.67  \\
$l_{6}$ &             Perturbation &           -2.06  \\
$l_{7}$ &             Perturbation &           -0.24  \\
$l_{8}$ &             Perturbation &           -0.60  \\

\end{tabular}
\tablefoot{The $l_{i}$ parameters are the coefficients of each $i$-th grade Legendre polynomial in the basis of Equation \ref{eq:pert}.}
\label{tab:model_QE_parameters}
\end{table}

\subsection{Atmospheric parameters for LAST}\label{sec:atm_LAST}
 The only atmospheric parameter currently measured on each telescope mount of the LAST array simultaneously to observations and registered in the image header is the temperature T. The reported value is used for each image calibration and kept fixed during the model optimization. The simultaneous measurement of the surface pressure p will be included in the future in the data products, meanwhile, in this application we fix it at the average value of 965\,mbar (hPa) at the site location. For the Ozone column, measurements\footnote{\url{https://www.temis.nl/uvradiation/UVarchive/stations_uv.php}} from stations nearby the LAST site show that $u_{0}$ varies slowly with time, with an average value of $\sim$300 DU and $\sigma\,\sim$ 30 DU. Deviations in this range, correspond to a variation in the photometric zero point < 1 mmag. Therefore we choose to fix the value of $u_{0}$ to 300 DU in the model. Lastly, given the fact that the aerosol parameter $\alpha$ is highly correlated with the aerosol optical depth $\tau_{a500}$ and its small contributions to the zero-point variations (see Figure \ref{fig:deltamag_atm}), we fix it to the typical value of 0.6 indicated by \citet{2019SoEn..187..233G} for desert-like/rural conditions.
 As shown in Sec. \ref{sec:sensitivity_atm}, among all six atmospheric parameters, only the aerosol optical depth ($\tau_{a500}$) and the amount of precipitable water ($p_{w}$) can significantly contribute to the photometric error assuming a broadband transmission model. This is also the case for LAST data, which are acquired in clear filter mode. Therefore these two parameters are kept free in our model, and are fitted simultaneously. A summary of the values for the atmospheric parameters used in the calibration of LAST data, their correspondent atmospheric component and their value or status in the fit is presented in Table \ref{tab:model_parameters}.

\begin{table}[h!]
\centering
\caption{Parameters for the atmospheric components of the transmission.}
\begin{tabular}{lll}
\hline
        \textbf{Parameter Name} &                  \textbf{Component} &                \textbf{Value (LAST)}  \\
\hline
$p$ &             Rayleigh, H$_{2}$O, UMG &           965  \\
$\tau_{a500}$ &             Aerosol &           0.084 (free)  \\
$\alpha$ &             Aerosol &           0.6  \\
$u_{0}$ &             Ozone &           300  \\
$p_{w}$ &             H$_{2}$O &          1.0 (free)  \\
T &             UMG &           measured  \\

\end{tabular}
\tablefoot{The surface temperature has no predefined value since it is measured on site at each image registration, and the values of $\tau_{a500}$ and $p_{w}$ are the nominal starting values before fitting it during the model optimiziation. The value column shows the parameter nominal (or fixed) value for LAST.}
\label{tab:model_parameters}
\end{table}

\subsection{Sources with \Gaia XP spectra in the LAST fields}\label{sec:gaiaxpspectra}
An important element required for this approach is to have, in each image, a sufficient number of stellar objects with calibrated spectra. This condition has historically been very difficult to meet, since the coverage of spectroscopy campaigns is usually limited, and their photometric calibration is often poor. For our goal of calibrating each observation independently, it is important to have a large number of calibrators spread uniformly across the sky and across a wide range of magnitudes and colors.

In the \Gaia Data Release 3 (\Gaia-DR3), a sample of 220 million low-resolution spectra for sources with G $<$ 17.65 mag are available in both the BP and RP bands. These spectra cover the wavelength range from 330-1050 nm \citep{2023A&A...674A...2D,2023A&A...674A...3M}, they are widely distributed across the sky and their coverage of colors and magnitudes make this an excellent sample to be used in our method. 

\begin{figure}
\includegraphics[width=\linewidth]{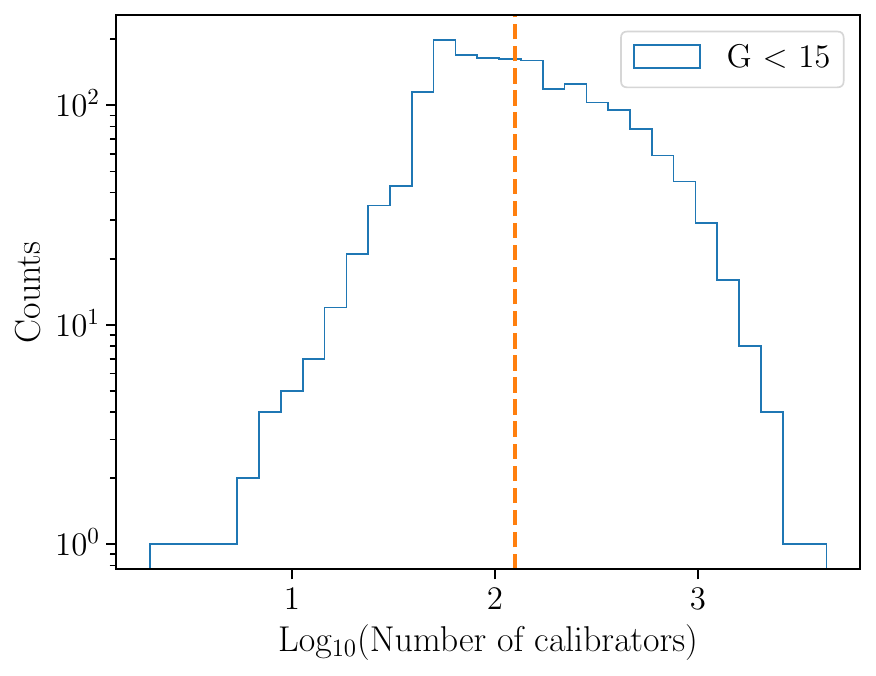}
\caption{Average number of calibrators with \Gaia XP spectrum and G $<$ 15 in area of $\approx$ 0.3 sq. deg. The vertical dashed orange line marks the median of the distribution ($\sim$130). 
 \label{fig:number_calibrators}}
\end{figure}

For the application on data from LAST, we generally use the sample of $\sim$35 million externally-calibrated BP and RP (named as XP in \Gaia products) spectra with G $\leq$ 15 and pre-sampled in the range 336-1020 nm. In the selection process of sources within each sub-image, we select only those with a probability\footnote{classprob\_dsc\_combmod\_star parameter in \Gaia products.} of being stellar objects $>$ 90$\%$. Figure \ref{fig:number_calibrators} shows the average number of available calibrators per sub-image for each field observed in the all-sky survey footprint of LAST. The median average number of calibrators in each sub-image is $\sim$130, and 4$\%$ of the fields have, on average, $<$30 calibrators. When the number of calibrators is less than 30, we extend the magnitude threshold for calibrators to G $<$ 16, and we retrieve raw spectra for these additional calibrators. When selecting from this larger sample, we externally calibrate the raw spectra with the GaiaXPy\footnote{\url{https://gaia-dpci.github.io/GaiaXPy-website/},\url{10.5281/zenodo.8239995}} software. We then resample them on an evenly-spaced wavelength array with $\Delta\lambda=$ 0.2 nm, to match the sampling of the pre-calibrated spectra. The spectra of fainter stars are typically of lower quality.

\begin{figure}
\includegraphics[width=\linewidth]{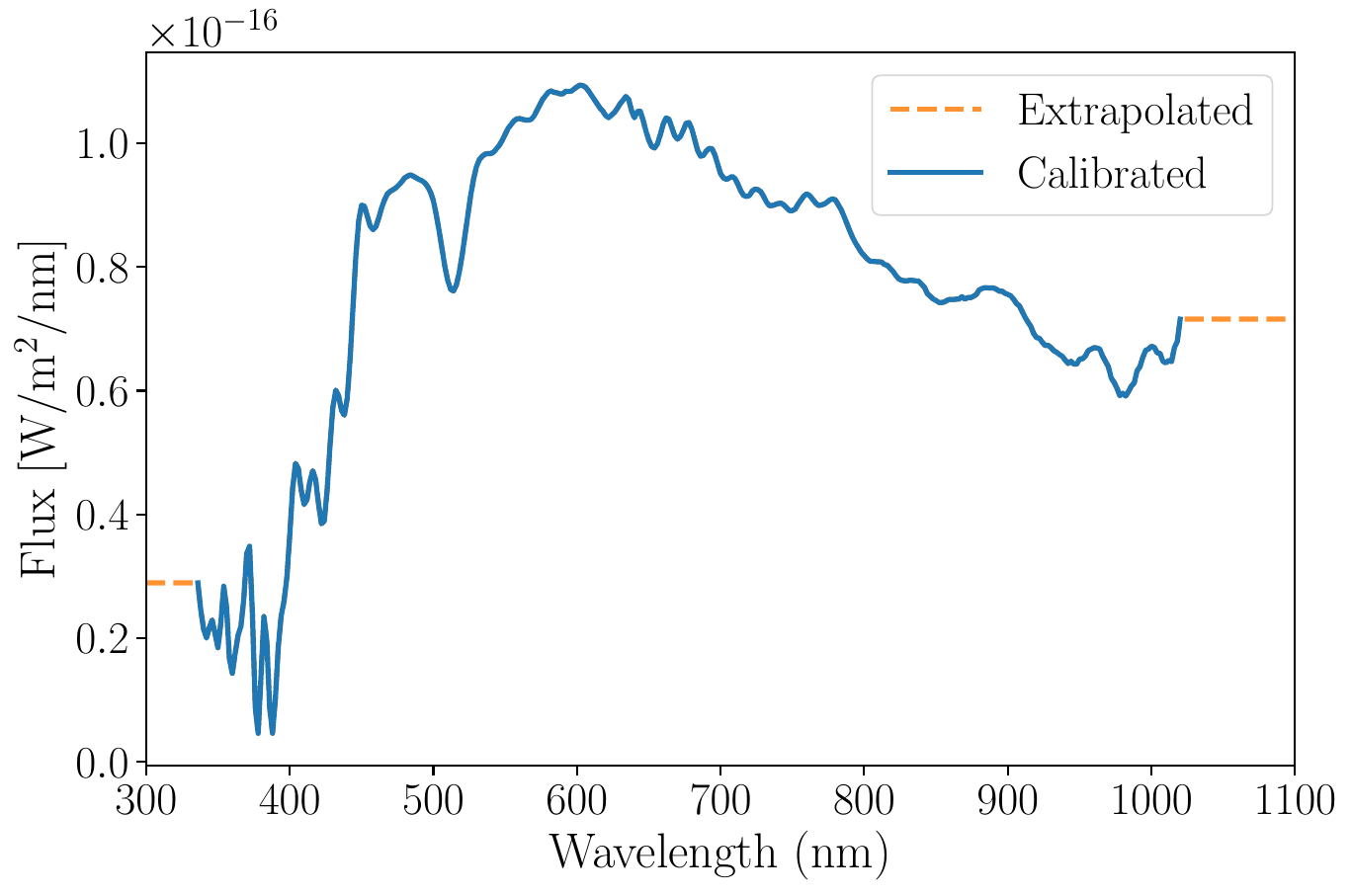}
\caption{Example of externally calibrated \Gaia spectrum (solid blue line) and its extrapolation (dashed orange line) in the 300-1100 nm range.
\label{fig:spectrum_extrapolation}}
\end{figure}

As shown in Figure \ref{fig:ota_transmission}, the instrumental transmission for LAST has a percent-level sensitivity in the ranges 300-336 nm and 1020-1100 nm. Therefore, we extrapolate the spectra outside the nominal wavelength range, assigning a constant flux value to F($\lambda$) equal to the first (last) calibrated value in the nominal wavelength range. An example of this extrapolation is shown in Figure \ref{fig:spectrum_extrapolation}, where the calibrated spectrum is shown as a continuous line and the extrapolated one as a dashed line. \\

\subsection{Application}\label{sec:application_real_data}
For the absolute calibration of LAST data, the model has a total of fourteen free parameters. These are divided as follows: one overall normalization ($C$, Equation \ref{eq:residuals_w_fieldcorr}), a set of three parameters in the transmission model ($\hat{\varrho}$, Equation \ref{eq:residuals_w_fieldcorr}), and a set of ten parameters in the polynomial field correction term ($\hat{k}$, Equation \ref{eq:residuals_w_fieldcorr}). $\hat{\varrho}$ contains the center of the QE model ($\mu$) and two atmospheric parameters ($p_{w}$ and $\tau_{a500}$). The last set of parameters $\hat{k}$ are the coefficient of the polynomial field correction terms (Equation \ref{eq:polynomial_field_corr}) for which we use a basis of Chebyshev polynomials up to the 4th with $n_{x} = \,$0,1 for $n_{y} \leq \,$4 and $n_{y} = \,$0,1 for $n_{x} \leq \,$4. The instrumental coordinates $x,y$ ranging from 0 to 1726 pix in each sub-image are transformed into the range of -1 and 1.

 For the model optimization, a particular attention goes in the estimation of the errors to weight the residuals in Equation \ref{eq:residuals_w_fieldcorr}. Since the uncertainty on the \Gaia spectra (typically on the order of 1--10\% per wavelength bin) dominates the error budget on the photometry of a single source, we estimate the photometric error derived by error propagation of the error in \Gaia spectra for each bin, multiplied with a nominal model of the LAST transmission, to obtain an error on the counts rate.
 
The model is fitted independently on each of the 24 sub-divisions of each image (i.e., in a 0.55 deg $\times$ 0.55 deg field of view). The fit is performed in two main steps: firstly, the normalization $C$, the center $\mu$ of the QE model and the ten $\hat{k}$ coefficients of the polynomial field correction are fitted simultaneously. In the second step, these parameters are fixed to the best-fit values and the two atmospheric parameters are freed and fit simultaneously. For each step, 3 iterations are performed and a 3$\sigma$ clipping is applied in order to remove stellar calibrators that have large residuals. Given the loose criteria in the selection of the stellar calibrators, this procedure helps with removing variable sources and problematic calibrators. Depending on the field, typically a fraction between 10-50\% of calibrators are clipped during the iterations, keeping a minimum value of 30 calibrators. The model residuals are minimized using the Levenberg–Marquardt algorithm for least-square fitting \citep{levenberg1944,newville_2015_11813}. 

Once obtained the best-fit model for the system transmission, the flux measurements of the sources in the sub-images are converted into magnitudes in the AB system \citep{1974ApJS...27...21O,1983ApJ...266..713O} calculating the zero-point using a flat-spectrum source with F($\nu$) = 3631 Jy and following the prescriptions from \citealt{2012PASP..124..140B}.

\section{Results based on LAST data}\label{sec:results}
\begin{figure*}[h!]
\centering
\includegraphics[width=17cm]{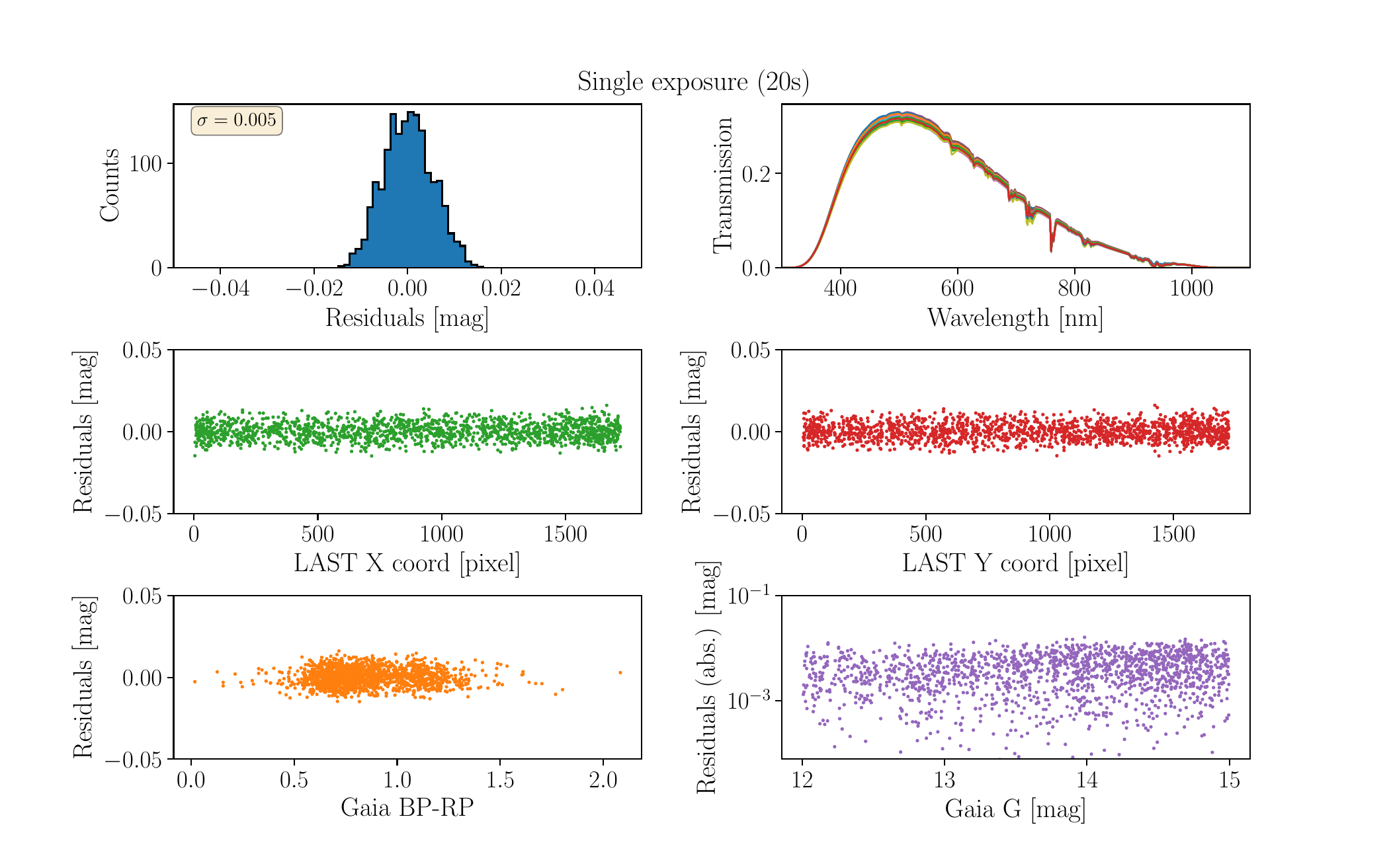}\\
\vspace{-1cm}
\includegraphics[width=17cm]{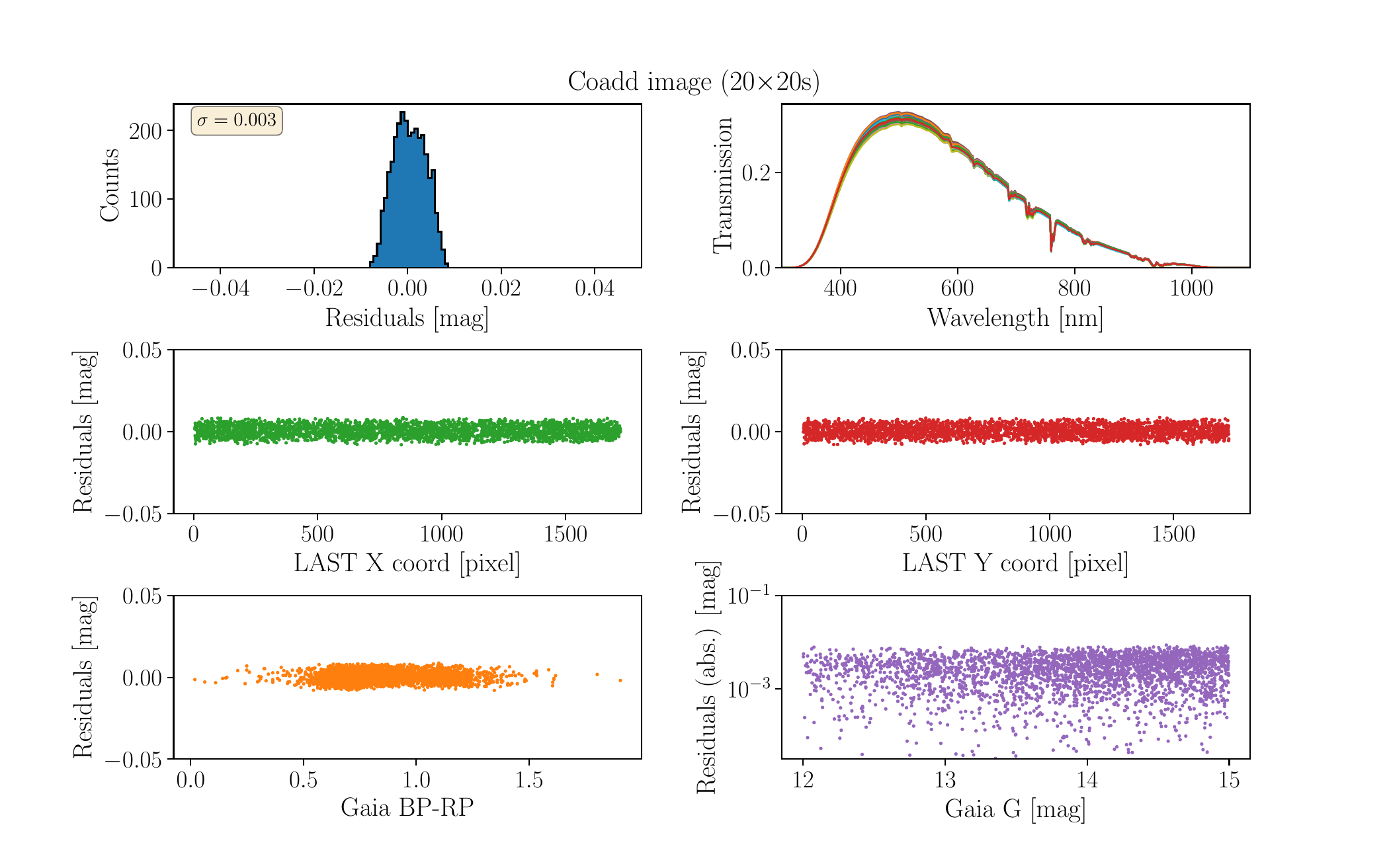}
\caption{Summary of residual distributions for a single 20 s exposure (top) and for 20$\times$20 s coadded exposures (bottom). The top-left panel shows the overall distribution of the residuals, with the best-fit standard deviation assuming a Gaussian shape. The top right panel shows the 24 transmissions of each sub-image of a single image, fitted independently. The middle panels show the residuals (for all sub-images) as a function of the instrumental coordinates of the camera sensor. The bottom panels show the residuals as a function of the Gaia color (left) and the \Gaia G magnitude (right), respectively. 
\label{fig:fullimage_diagnostic}}
\end{figure*}

We apply the calibration procedure described in the previous section to a sample of LAST data. We compare the results of both single 20 s exposures and 20x20 s coadded images, to compare the calibration with increased overall signal-to-noise ratio in the image. The behavior of the calibration residuals throughout the image is summarized in Figure \ref{fig:fullimage_diagnostic}, were we show diagnostic plots for a LAST image (24 sub-images fitted independently) for a single 20 s exposure (top block) and for an image resulting from the coaddition of 20$\times$20 s exposures. For the single 20 s exposure, residual distributions (top left) are very similar across all the sub-images, with an overall Gaussian-like distribution centered around 0 with $\sigma $= 0.5$\%$ in the example shown. The standard deviation of the residual distribution varies slightly depending on the field and observing conditions, but we find it to be generally $< 1\%$. The top right panel of Figure \ref{fig:fullimage_diagnostic} shows the fitted transmissions of each sub-image. It can be noticed that the best-fit solutions are very similar to each other, with small position-dependent differences across the image.
The mid panels of Figure \ref{fig:fullimage_diagnostic} show the model residual scatter as a function of the instrumental coordinates of the sensors on the X (green dots, mid-left panel) and Y (red dots, mid-right panel) axes for all sub-images. No correlated trends are observed, proving the effectiveness of the polynomial field-correction term in the model. 
The bottom-left panel shows the residual scatter as a function of the \Gaia BP-RP color of the calibrators, and no correlated trends are shown (see Appendix \ref{app:appendix_b} for an example of color-dependent trends). 
The bottom-right panel shows the residual scatter as a function of the mean \Gaia G magnitude. This shows a slow average increase towards higher magnitudes, as expected from the degrading of the quality of the spectra and photometry.

The bottom block of Figure \ref{fig:fullimage_diagnostic} shows the same residuals distributions this time for a $20\times20$\,s coadded image. As expected, the increase in the overall signal-to-noise ratio leads to an improvement of the calibration accuracy.

In Appendix \ref{app:appendix_c}, we discuss the effect of the polynomial corrections introduced in Sec. \ref{sec:method} on the residuals across the sub-images and on the resulting spatial distribution of the photometric zero point.

\subsection{Zero-point error from non-parametric bootstrap}
\label{sec:bootstrap}
\begin{figure}[h]
\includegraphics[width=\linewidth]{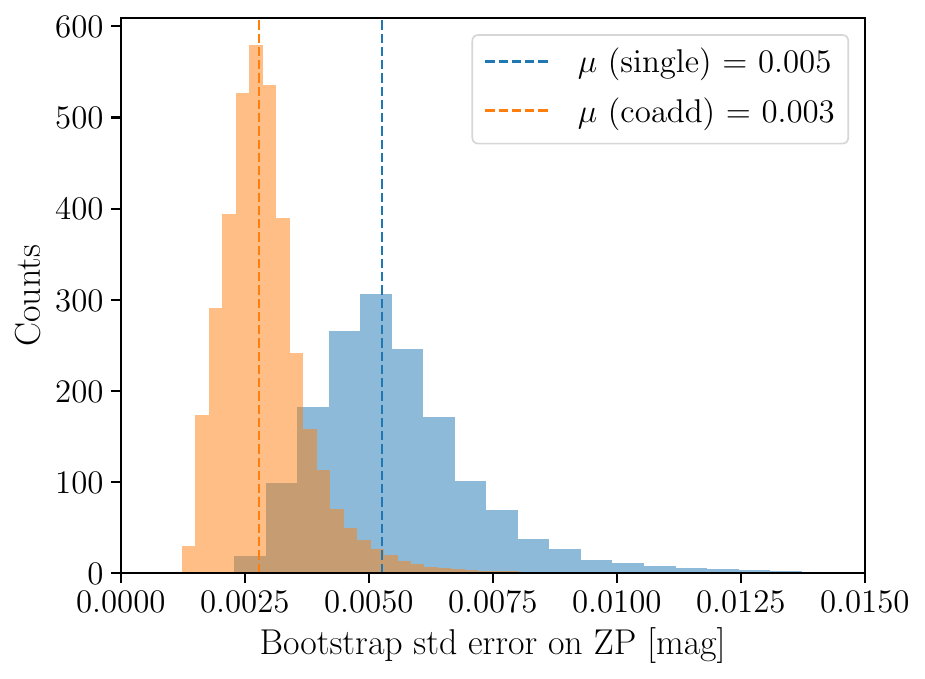}
\caption{Distributions of the bootstrap standard error on the AB zero point for all LAST sub-images of a single exposure (blue) and for 20x20 s coadded images (orange). The dashed lines mark the medians of each distribution.
\label{fig:bs_2d_map_subimage}}
\end{figure}

Another way to quantify the accuracy of the model is to estimate the standard error on the zero-point via a non-parametric bootstrap process. To do so, we follow the procedure in \citealt{efron1994introduction}, and realize N calibrator sets for each sub-image by drawing randomly with repetitions from the original set of calibrators. Each random set of calibrators has the same size as the original one and it is used to derive a new set of best-fit parameters for the transmission model. After computing the coordinate-dependent transmission for all N bootstrap representations, the bootstrap standard error on the zero point can be calculated for each point in the image (in our case, we create a grid with 30 pixel spacing). We show in Figure \ref{fig:bs_2d_map_subimage} the distributions of the standard error evaluated for all sub-images in a single 20 s exposure (blue) and for all sub-images in a 20x20 s coadded image (orange). The distribution for the single 20 s exposure has a median value of $\sim5$ mmag, with values ranging from $\sim2$ mmag and a longer tail ranging up to $\sim12$ mmag. This tail is likely due to the degrading at the very edges of each sub-image. The median value of $\sim5$ mmag is a good representation of the zero-point error we get when we apply this test to many observations taken with different LAST units. The orange distribution shows the error on the zero point for each grid point in a 20x20 s coadded image. In this case, the median of the distribution is lower at $\sim3$ mmag, with values ranging from $\sim1$ mmag to $\sim7$ mmag.

\subsection{Stability over time}\label{sec:temporal_stability}
\begin{figure}[ht!]
\includegraphics[width=\linewidth]{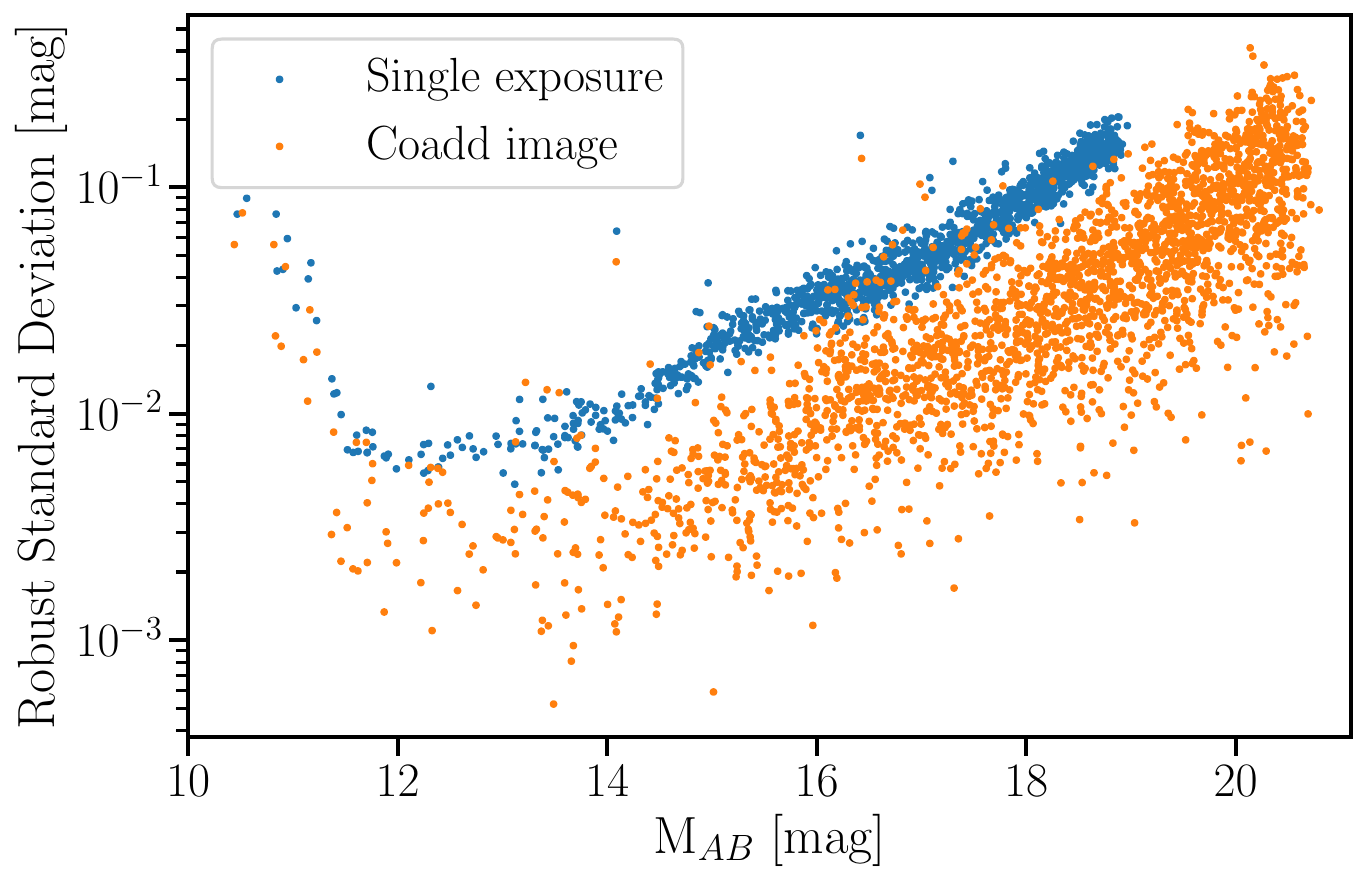}
\caption{Robust standard deviation ($\sigma^{\prime}$) as a function of the mean AB magnitude for all sources in a LAST sub-image using data from 20 single images per night (blue points) and one 20$\times$20 s coadded image per night (orange points) for a total of 8 nights. The points are calculated for both aperture and psf photometry for each source, then the one with smaller $\sigma^\prime$ is chosen. 
\label{fig:rms_plot_singlefield_survey}}
\end{figure}

In order to quantify the stability over time of the photometric calibration, we consider a series of 20$\times$20 s observations per night of the same field, taken for a total of 8 nights between 3--12 March 2024. Each batch of 20 images is observed continuously, and in LAST defines a standard visit for which coadded images are created. After calibrating each sub-image individually, we consider all sources within a single sub-image. We show in Figure \ref{fig:rms_plot_singlefield_survey} the robust standard deviation ($\sigma^\prime$)\footnote{\url{https://docs.astropy.org/en/latest/api/astropy.stats.mad_std.html}} as a function of the average AB magnitude for all sources (blue points) within the sub-image with at least 80 detections in the whole sample of observations. The points are calculated for both aperture and PSF photometry for each source, then the one with the smaller $\sigma^\prime$ is chosen. Typically, sources with M$_{AB}$ $\leq$ 16 have better $\sigma^\prime$ with aperture photometry. The scatter reaches a 5-7 mmag level towards the bright end of the distribution. At M$_{AB}$ $<$ 12 the sources are saturated in the image. The orange points show the $\sigma^\prime$ versus M$_{AB}$ for the same sources in coadded images (20$\times$20 s) for the same observations. Compared to the blue points, the distribution spreads on a wider range of $\sigma^\prime$ because it is calculated on only 8 images resulting from the coaddition of the 160 original images into groups of 20. However, the improvement in stability from these images is clear, as the $\sigma^\prime$ reaches values as low as a few mmag on the bright end, and stay below 1$\%$ up to M$_{AB} \sim$16.

It is worth noting that secondary color terms might be present in the calibration, mostly coming from the reddening of sources at high air masses. In the application to LAST data we find these secondary color terms are similar between different telescopes and stable in observations form multiple nights. The color term is linear with typical values of 10-50 mmag/airmass, depending on the color of the source. This correction can be easily calculated after the main photometric calibration, by having two or more observations at different airmass.    

\subsection{Nightly evolution of the transmission}
\begin{figure}[ht!]
\includegraphics[width=\linewidth]{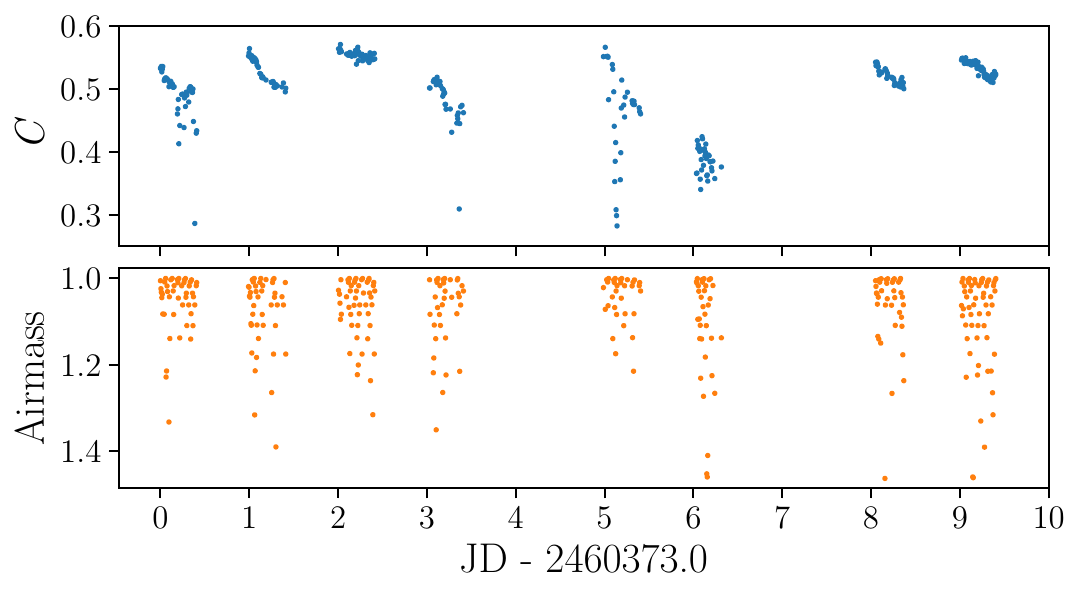}
\caption{Upper panel: Overall normalization parameter for several sky-survey observations obtained during the same nights of the temporal stability study in Sec. \ref{sec:temporal_stability}. Each blue point represents the parameter $C$ (from Eq. \ref{eq:residuals_w_fieldcorr}) for a coadded image. Lower panel: Corresponding airmass of each observation (orange points). 
\label{fig:nightly_evolution_transmission}}
\end{figure}

It is interesting to have a glimpse at how the transmission varies during the night and between different nights. To do so, in the upper panel of Figure \ref{fig:nightly_evolution_transmission} we show the evolution of the overall normalization parameter $C$ from Eq. \ref{eq:residuals_w_fieldcorr}, which dominates the amplitude of the transmission as function of time. We show data from the same nights of the temporal stability study in Sec. \ref{sec:temporal_stability}, but this time we use observations from several fields observed throughout each night during the sky survey operations. We see, as expected in the period considered (March 2024), a worsening of the conditions of the site during the night, mostly due to the increased humidity and formation of thin layers of clouds. During the fifth night we see more dramatic variations of conditions with gradual drops of $C$ from $\sim$0.55 to $\leq$0.3 within 1-2 hours and then subsequent gradual improvements. In the lower panel we show the corresponding airmass of each observations, to remark that these trends are not correlated with airmass changes, but can be interpreted as a broad metric of the observing conditions. However, we stress on the fact that the parameter $C$ is an overall normalization of the model, therefore its long-term variation can be due also to instrumental effects. For example, we notice average increases of this parameter up to 10-20$\%$ after regular cleaning operations of the telescope corrector plates.   

\section{Software}\label{sec:software}
The implementation of this method is publicly available on Github\footnote{\url{https://github.com/simonegarrappa/transmission_fitter.git}} in its python implementation. A MATLAB version is under development and it will be included in the AstroPack/MAATv2\footnote{\url{https://github.com/EranOfek/AstroPack}} Astronomy $\&$ Astrophysics Toolbox \citep{AstroPack}. 
For the atmospheric components, our code is based on the parameterizations and implementations from the SMARTS v2.9.8 model \citep{2019SoEn..187..233G} and its python implementation from \citealt{RUIZARIAS2022112302}. \Gaia spectra are retrieved either through Astroquery \citep{2019AJ....157...98G} or a fast local catalog matching through {\tt catsHTM} \citep{2018PASP..130g5002S}. Lastly, the model optimization is performed using the {\tt lmfit} package \citep{newville_2015_11813}.

At the date of this preprint, the software contains mainly libraries to apply the model on LAST data. Libraries to handle data products from other instruments will be added in the futures, as well as more generic base models to make the code more versatile to any filter shape. 
In Appendix \ref{app:header_keys} we suggest formats to report the calibration parameter in FITS headers.

\section{Summary}\label{sec:summary}
We presented a novel approach to the absolute photometric calibration of ground-based telescopes. This method consists of deriving the transmission of the system and aims to break the 1$\%$ barrier of accuracy that is typical of the current state-of-the-art absolute calibrations.
Furthermore, it is designed to over come the problem of comparing photometric data obtained using different telescopes, or the same telescope at different atmospheric conditions. In this case, a rough spectral model of the source can be used to convert observations taken with different instruments to the same passband. A key point of this method is that the transmission model can be fit independently for each image, given a large enough number of calibrators in the field of view.

From an application to LAST data we showed that, by fitting an appropriate subset of the instrumental and atmospheric parameters for each image, we achieve a median accuracy of the photometric zero point that ranges between 3-5 mmag, depending on the total exposure of the image. Moreover, stability of the zero-point accuracy is observed over long time scales. The improvement observed in 20$\times$20 s coadded images is of a factor $\sim$1.5--2. We notice that this is lower than a factor of $\sqrt{n}\sim$4.5, likely because we are reaching the systematic error level in \Gaia photometry of about 3 mmag. Lastly, the spatial photometric corrections that are simultaneously fitted with the transmission provide a uniform accuracy across the whole image and do not show dependencies on the spectral color of the calibrators.

Thanks to its high versatility, this approach can be applied to any ground-based facility that has a large enough field of view, with the appropriate changes on the model assumptions for the instrumental response. The image-by-image calibration approach also allows for an efficient calibration of existing and future data from other telescopes. This approach is especially useful for wide field surveys like LAST, for which the high multiplicity of the instrument array requires an accurate, image-wise calibration approach. Since one of the basic requirements for this approach is to have a fair number of \Gaia spectra, this approach is more efficient for wide-field instruments. However, if the current sample of $\sim$220 million spectra measured by \Gaia will increase in the future, it will increase the efficiency of this method also for instruments with a smaller field of view.

The method presented in this work will be adopted as the official absolute calibration for the LAST telescopes and it will be expanded to be applied also to monitor the evolution of the system throughput in space telescopes (e.g., ULTRASAT \citealt{2024ApJ...964...74S}). With space telescopes, dedicated campaigns observing a sample of calibrators at regular time intervals will be sufficient to fit the transmission from multiple images.

\begin{acknowledgements}
We would like to thank the anonymous referee for useful comments on the manuscript. We would also like to thank G. Altavilla for useful comments on the preprint.
S.G. is grateful for the support of the Koshland Family Foundation.
E.O.O. is grateful for the support of
grants from the 
Willner Family Leadership Institute,
André Deloro Institute,
Paul and Tina Gardner,
The Norman E Alexander Family M Foundation ULTRASAT Data Center Fund,
Israel Science Foundation,
Israeli Ministry of Science,
Minerva,
BSF, BSF-transformative, NSF-BSF,
Israel Council for Higher Education (VATAT),
Sagol Weizmann-MIT,
Yeda-Sela, and the
Rosa and Emilio Segre Research Award.
This research is supported by the Israeli Council for Higher Education (CHE) via the Weizmann Data Science Research Center, and by a research grant from the Estate of Harry Schutzman.
\end{acknowledgements}

\bibliographystyle{aa} 
\bibliography{biblio.bib} 

\begin{appendix}

\section{Sensitivity to variations of the transmission average wavelength}\label{app:appendix_a}

\begin{figure}[ht!]
\includegraphics[width=\linewidth]{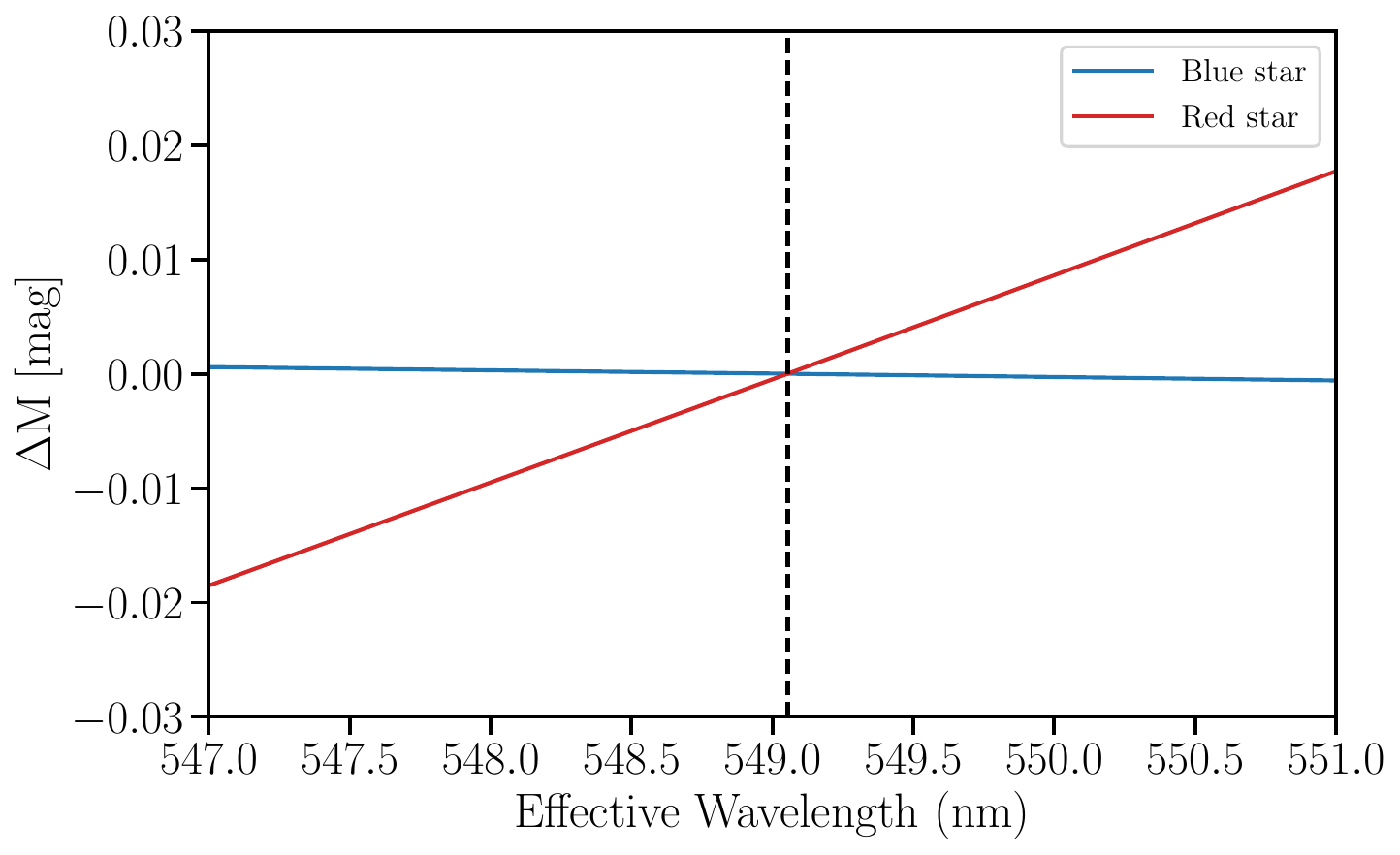}
\caption{Comparison of the magnitude variation as a function of the effective wavelength of the transmission between a blue and a red star.
\label{fig:deltamag_vs_avwvl}}
\end{figure}

We provide here an example to compare the sensitivity of small variations in the average wavelength of a passband. We consider a nominal transmission for the LAST system, described in Sec. \ref{sec:instr_transmission_LAST} and we vary its average wavelength ($\lambda_{0}$, \citealt{1986HiA.....7..833K}) defined as

\begin{equation}
    \lambda_{0} = \frac{\int T(\lambda)\lambda d\lambda}{\int T(\lambda) d\lambda}
\end{equation}
between 547.0 nm and 551.0 nm. We select two \Gaia spectra from a blue and a red star, respectively, and we calculate the magnitude of the source as observed in each passband assuming a zero point of 25 mag. Figure \ref{fig:deltamag_vs_avwvl} shows the magnitude difference in each transmission calculated starting from the nominal passband with $\lambda_{0}$ = 549.05 nm. We see how these small variations in the transmission have a small effect (< 10$^{-3}$) for the blue star, while they reflect in a difference of a few \% for the red star. This example shows how even small changes in the system transmission can affect dramatically the photometric stability.

\section{The importance of fitting the transmission center $\mu$}\label{app:appendix_b}
\begin{figure}[ht!]
\includegraphics[width=\linewidth]{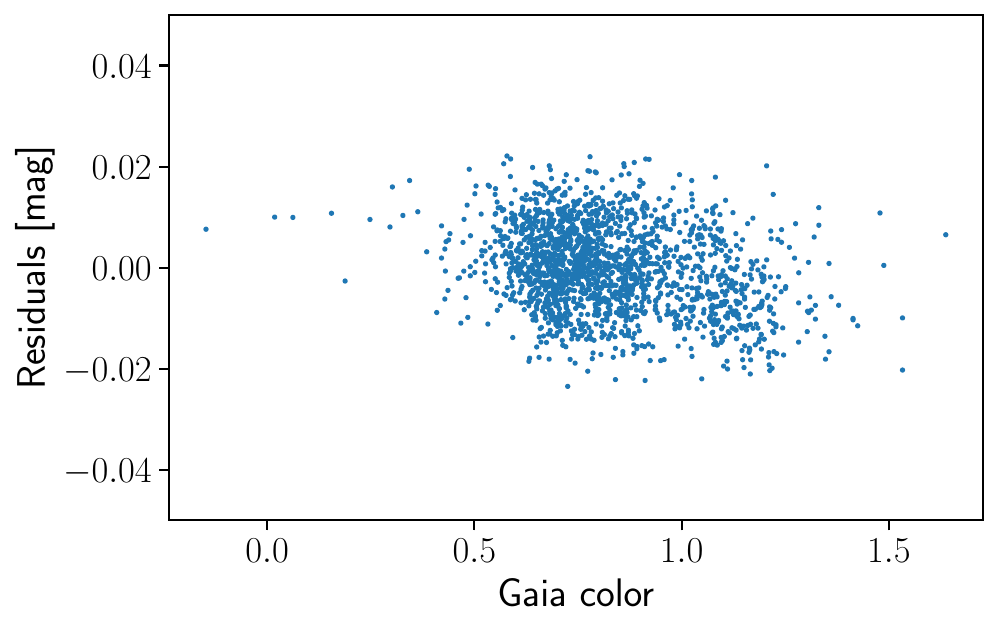}
\caption{Residuals vs \Gaia BP-RP color for a fixed-shape QE analytical model in the calibration procedure.
\label{fig:colordependency}}
\end{figure}

Here we explore the effect of fitting the shape of the analytical model used for the detector quantum efficiency. We run the calibration on the single 20 s exposure calibrated in Sec. \ref{sec:results} by keeping fixed the parameter $\mu$ of the skewed Gaussian model at $\mu\,=$ 390.1 (the best-fit value on the data from \citealt{2023PASP..135f5001O}). Figure \ref{fig:colordependency} shows the calibration residuals for all sources with a \Gaia counterpart as function of their color. Differently to the behavior in Figure \ref{fig:fullimage_diagnostic}, in this case a correlation with the \Gaia BP-RP color can be seen in the distribution. For this particular case, the correlation coefficient is $\sim-0.3$, while stronger correlations can be generally found in individual subframes throughout the image.

\section{The importance of the zero-point field corrections and the effect of gray structures.}\label{app:appendix_c}
    
\begin{figure}[ht!]
\includegraphics[width=\linewidth]{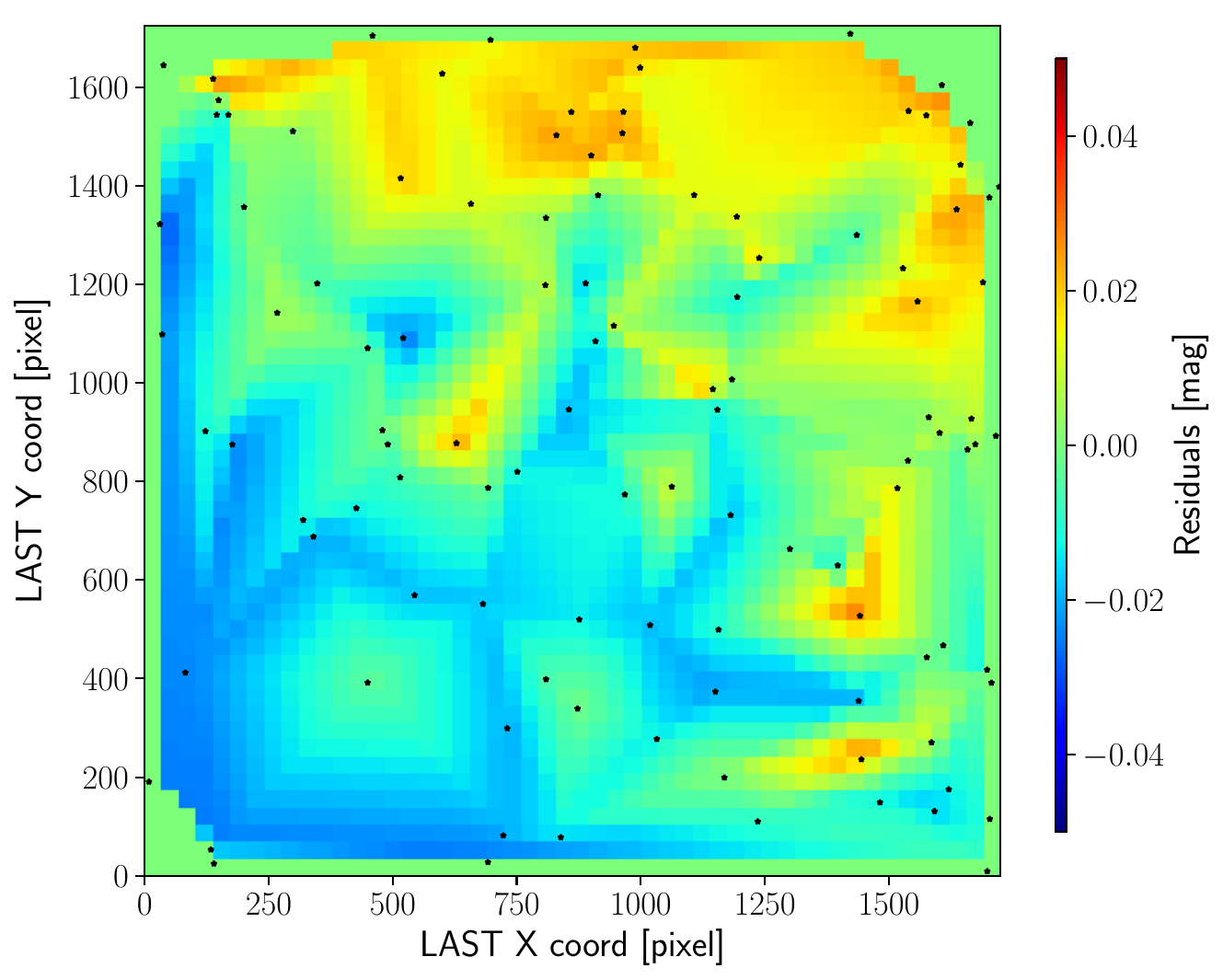}
\caption{Residuals as function of the detector coordinates in a LAST sub-image. In this case, the field correction terms are removed from the model. The residuals are calculated for each calibrator star (black dots) and linearly interpolated in a grid.
\label{fig:noFieldCorr}}
\end{figure}

We repeat the same exercise but this time we remove the polynomial field correction terms from the model. We show in Figure \ref{fig:noFieldCorr} an example of calibration residuals in a LAST sub-image when the field correction terms from Equation \ref{eq:residuals_w_fieldcorr} are removed. A strong gradient of the residuals from the bottom left to the top right corner of the frame can be distinguished in the colormap.     

\begin{figure*}[]
\includegraphics[width=0.5\textwidth]{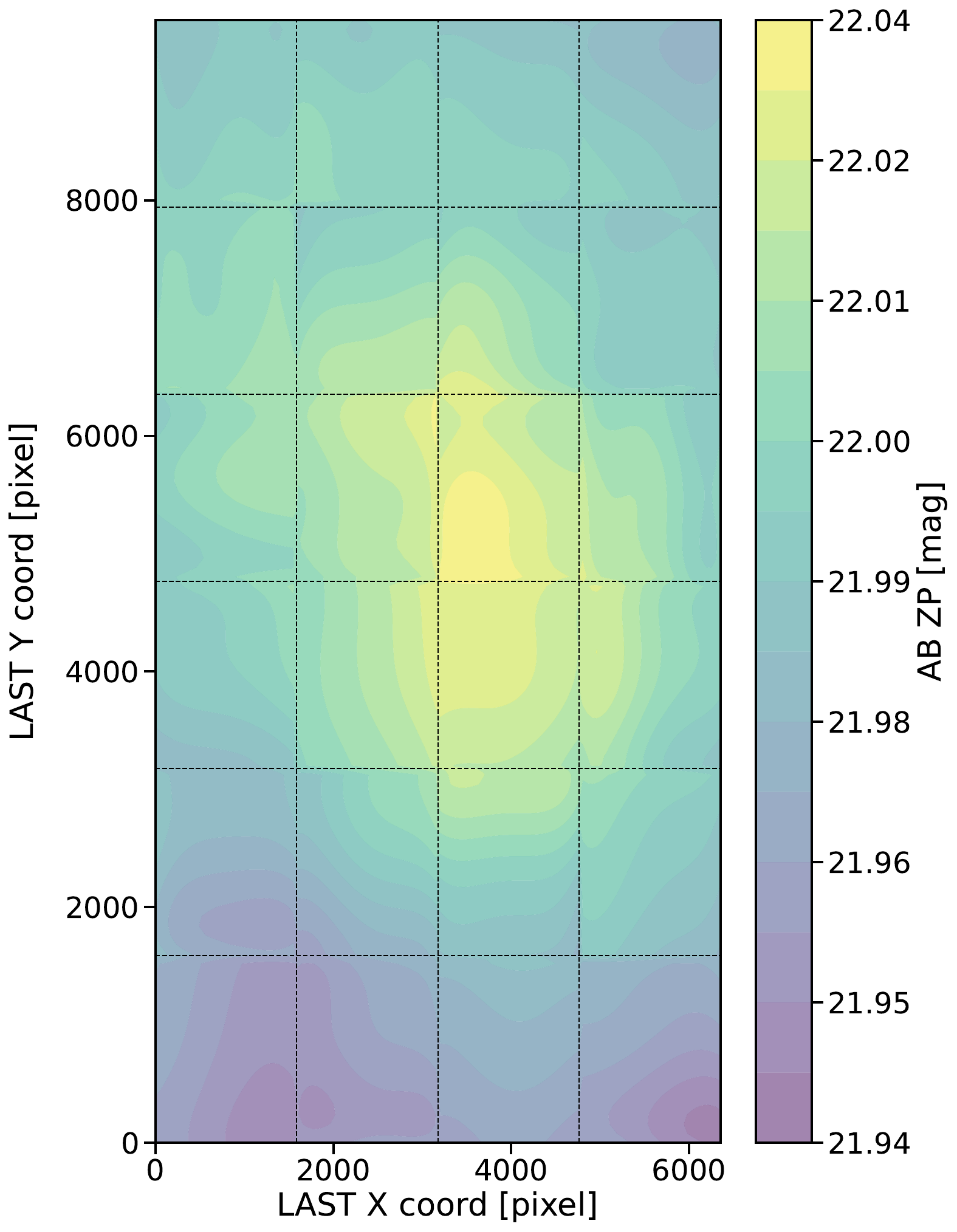}
\includegraphics[width=0.5\textwidth]{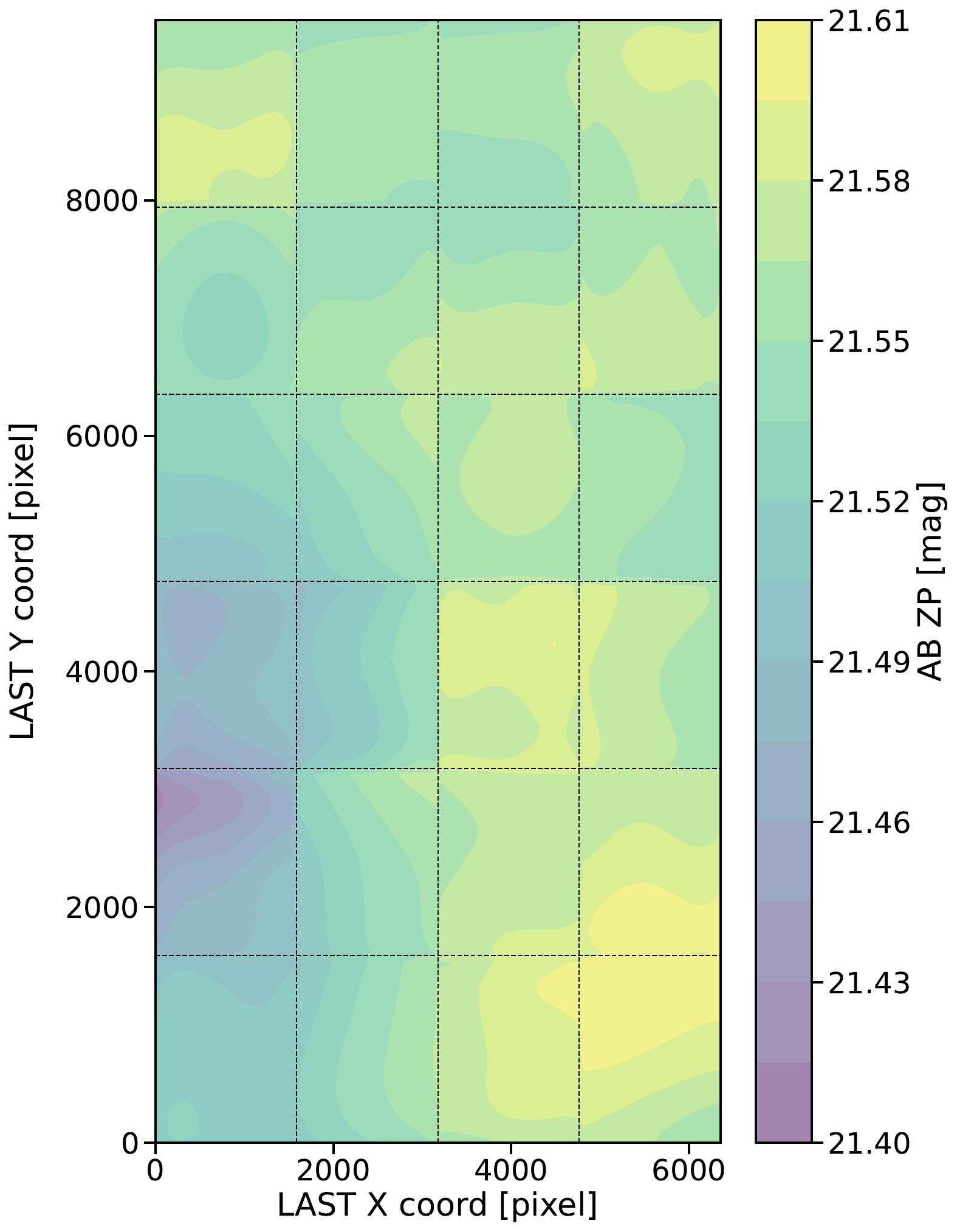}
\caption{Zero-point 2-D distribution for an image with low presence of cloud structures (left) and one with significant presence of cloud structures (right). The dashed lines mark the sub-images, which are calibrated independently.  
\label{fig:gray_structures}}
\end{figure*}

When present, these gradients are observed with different patterns that only depend on the specific image. Therefore these might be related to residual non-uniformity after the flat-field correction or some temporary instrumental gradient on the camera sensor. We observe non-uniformity in the residuals up to a $\sim$10$\%$ level in some of the most extreme cases.
As already mentioned in Sec. \ref{sec:method}, the polynomial field corrections (together with the overall model normalization), are important to correct for wavelength-independent (gray) extinction caused by clouds. However, it is hard in the current formulation of the method to disentangle the contribution of gray extinction and instrumental non-uniformity. Moreover, other effects might contribute significantly (e.g., PSF variation across the field of view). Nevertheless, from the spatial distribution of the zero-point, it is possible to distinguish cases in which instrumental non-uniformities are dominant, to cases where the field is dominated by gray structures. In the left panel of Figure \ref{fig:gray_structures} we show an image in which the zero point spatial distribution is dominated by the typical vignetting pattern (likely residuals from flat-field corrections). In the right panel, we show an image in which the gray structures are likely dominant, and affect significantly the spatial distribution of the zero-point values. It is interesting to notice how the structures are consistent across multiple sub-images, which are calibrated independently. This supports the robustness and consistency of our approach. This method, utilizing a large number of LAST telescopes observing a continuous patch of the sky, may be used to measure the power spectrum of the gray extinction. This may be investigated in a future work.

\section{Airmass parameterizations for atmospheric components}\label{app:airmass_pars}
Here we report in Table \ref{tab:airmass_parameters} the parameterizations from the SMARTS v2.9.8 model \citep{2019SoEn..187..233G} used to calculate the effective airmass of each atmospheric component. The details on which component is included for each atmospheric transmission are described in Sec. \ref{sec:atmospheric_transmission}.

\begin{table}
\centering
\caption{Parameters used to calculate the effective airmass of each atmospheric component.}    
\begin{tabular}{lrrrr}
\hline
\textbf{Component} & \textbf{$\eta_{0}$} & \textbf{$\eta_{1}$} & \textbf{$\eta_{2}$} & \textbf{$\eta_{3}$} \\
\hline
Rayleigh & 0.48353 & 0.095846 & 96.741 & -1.754 \\
Aerosol & 0.16851 & 0.18198 & 95.318 & -1.9542 \\
$O_3$ & 1.0651 & 0.6379 & 101.8 & -2.2694 \\
$H_{2}O$ & 0.10648 & 0.11423 & 93.781 & -1.9203 \\
$O_{2}$ (UMG) & 0.65779 & 0.064713 & 96.974 & -1.8084 \\
$CH_{4}$ (UMG) & 0.49381 & 0.35569 & 98.23 & -2.1616 \\
$CO$ (UMG) & 0.505 & 0.063191 & 95.899 & -1.917 \\
$N_{2}O$ (UMG) & 0.61696 & 0.060787 & 96.632 & -1.8279 \\
$CO_{2}$ (UMG) & 0.65786 & 0.064688 & 96.974 & -1.8083 \\
$N_{2}$ (UMG) & 0.38155 & 0.000089 & 95.195 & -1.8053 \\
$HNO_{3}$ (UMG) & 1.044 & 0.78456 & 103.15 & -2.4794 \\
$NO_{2}$ (UMG) & 1.1212 & 1.6132 & 111.55 & -3.2629 \\
$NO$ (UMG) & 0.77738 & 0.11075 & 100.34 & -1.5794 \\
$SO_{2}$ (UMG) & 0.63454 & 0.00992 & 95.804 & -2.0573 \\
$NH_{3}$ (UMG) & 0.32101 & 0.010793 & 94.337 & -2.0548 \\
\hline
\end{tabular}
\tablefoot{Parameterizations are from the SMARTS v2.9.8 model \citep{2019SoEn..187..233G}.}
\label{tab:airmass_parameters}
\end{table}

\section{Suggested standard for FITS header keywords}\label{app:header_keys}
Here we suggest a common scheme for FITS\footnote{\url{https://fits.gsfc.nasa.gov}} header keywords to report the values of the model parameters in each image. Assuming the more general form of the transmission model, about 30 parameters are needed (e.g., Tables \ref{tab:model_QE_parameters},\ref{tab:model_parameters}). 

In our scheme, each header keyword will appear in the format {\tt PT\_$\langle$PARNUM$\rangle$\_$\langle$TYPE$\rangle$}, where {\tt PT} defines parameters for the photometric transmission, {\tt $\langle$PARNUM$\rangle$} is a 3-digits running number to distinguish between different parameters, and {\tt $\langle$TYPE$\rangle$} can be:

\begin{itemize}
\item {\tt N}: for keywords storing the extended unique name for the parameter (e.g., ATM$\_$aerosol$\_$alpha);
\item {\tt V}: for keywords storing the parameter value;
\item {\tt E}: optional, for keywords storing the parameter error;
\item {\tt F}: for keywords storing a boolean flag associated with the parameter that indicates if the parameter was free in the fit (1) or fixed (0).
\end{itemize}
Each parameter keyword will then appear up to four times with a different {\tt $\langle$TYPE$\rangle$}. Note that we do not force a unique mapping between the name and {\tt $\langle$PARNUM$\rangle$}. In addition, we suggest special keywords for the whole photometric calibration that are listed in Table \ref{tab:special_keys}. 

\begin{table}
\centering
\caption{Special header keys for the photometric calibration.}    
\begin{tabular}{ll}
\hline
\textbf{KEY} & \textbf{Description} \\
\hline
{\tt PT\_RMS} & photometric error\\
{\tt PT\_CHI2} & $\chi^2$ of the final fit  \\
{\tt PT\_DOF} & degrees of freedom of the final fit \\
{\tt PT\_NCALIB} & Number of calibrators \\
{\tt PT\_AREF} & Atmospheric model (e.g., SMART v2.9.8)   \\
{\tt PT\_SREF} & Software (e.g., pyabscal v1.0.0)  \\
{\tt PT\_SPEC} & Reference for spectra (e.g., \Gaia-DR3)  \\
\hline
{\tt PT\_LMIN} & Minumum wavelength of the transmission  \\
{\tt PT\_LMAX} & Maximum wavelength of the transmission  \\
{\tt PT\_DEL} & $\Delta\lambda$ of sampling \\
{\tt PT\_LUN} & Wavelength units\\
{\tt PT\_QE} & File name containing the sampled QE\\
{\tt PT\_TRSM} & File name containing the sampled transmission\\

\hline
\end{tabular}
\label{tab:special_keys}
\end{table}

The user has the freedom to use custom names for parameters, but we want to stress on the importance to adopt the same convention in the name structure of the header keywords. We list examples for the header keywords used in this work in Table \ref{tab:naming_convention}.

\begin{table}
\centering
\caption{Example of header keywords for the parameters used in this work.}    
\begin{tabular}{lll}
\hline
\textbf{KEY} & \textbf{Content} &\textbf{Description} \\
\hline
{\tt PT\_$\langle$PARNUM$\rangle$\_N} & NORM & $C$ in Eq. \ref{eq:residuals_w_fieldcorr} \\
\hline
{\tt PT\_$\langle$PARNUM$\rangle$\_N} & QE\_SG\_amplitude & A in Tab. \ref{tab:model_QE_parameters} \\
{\tt PT\_$\langle$PARNUM$\rangle$\_N} & QE\_SG\_mu & $\mu$ in Tab. \ref{tab:model_QE_parameters} \\
{\tt PT\_$\langle$PARNUM$\rangle$\_N} & QE\_SG\_sigma & $\sigma$ in Tab. \ref{tab:model_QE_parameters} \\
{\tt PT\_$\langle$PARNUM$\rangle$\_N} & QE\_SG\_gamma & $\gamma$ in Tab. \ref{tab:model_QE_parameters} \\
{\tt PT\_$\langle$PARNUM$\rangle$\_N} & QE\_PERT\_L\_$j$ & $l_{j}$ in Tab. \ref{tab:model_QE_parameters}  \\
\hline
{\tt PT\_$\langle$PARNUM$\rangle$\_N} & ATM\_p & $p$ in Tab. \ref{tab:model_parameters} \\
{\tt PT\_$\langle$PARNUM$\rangle$\_N} & ATM\_aerosol\_tau & $\tau_{a500}$ in Tab. \ref{tab:model_parameters} \\
{\tt PT\_$\langle$PARNUM$\rangle$\_N} & ATM\_aerosol\_alpha & $\alpha$ in Tab. \ref{tab:model_parameters} \\
{\tt PT\_$\langle$PARNUM$\rangle$\_N} & ATM\_ozone & $u_{0}$ in Tab. \ref{tab:model_parameters} \\
{\tt PT\_$\langle$PARNUM$\rangle$\_N} & ATM\_water\_pw & $p_{w}$ in Tab. \ref{tab:model_parameters} \\
{\tt PT\_$\langle$PARNUM$\rangle$\_N} & ATM\_temp & $T$ in Tab. \ref{tab:model_parameters} \\
{\tt PT\_$\langle$PARNUM$\rangle$\_N} & FC\_k\_$i$\_$j$ & $k_{ij}$ in Eq. \ref{eq:polynomial_field_corr} \\

\hline
\end{tabular}
\label{tab:naming_convention}
\end{table}

We suggest also a keyword scheme to store the transmission after sampling it in a number of wavelength bins N$_{\lambda}$ and to report the value of the transmission in each wavelength bin. This will also require to report four additional special header keywords (Table \ref{tab:special_keys}) for the minimum wavelength ({\tt PT\_LMIN}), maximum wavelength ({\tt PT\_LMAX}) sampling width ({\tt PT\_LDEL}) and wavelength units ({\tt PT\_LUN}). Alternatively, a sampled version of the QE or the full transmission can be stored in an external file which path is stored in the special keywords {\tt PT\_QE} and {\tt PT\_TRSM}. These special keywords are listed in Table \ref{tab:special_keys}. The transmission values are stored for each bin in three keywords:

\begin{itemize}

\item {\tt PT\_$\langle$BINNUM$\rangle$\_L} which contains the central wavelength of the bin; 

\item {\tt PT\_$\langle$BINNUM$\rangle$\_T} which contains the transmission value in the bin;

\item {\tt PT\_$\langle$BINNUM$\rangle$\_E} optional, which contains the transmission error in the bin.

\end{itemize}

Also here, $\langle$BINNUM$\rangle$ is a running 3-digits number. These keywords will therefore have the form, e.g., for bin 1: {\tt PT\_001\_L}, {\tt PT\_001\_T} and {\tt PT\_001\_E}.

Independently from the convention adopted for the header keywords, each source in the catalog results will have an additional column that is the AB zero-point calculated at the physical location of the source on the camera sensor.

\end{appendix}

\end{document}